\documentclass[twocolumn, trackchanges]{aastex631}
\usepackage{tabularx}
\usepackage[flushleft]{threeparttable}
\usepackage{amsmath}
\usepackage{graphicx}
\usepackage{epsf}
\usepackage{color}
\usepackage{pifont}
\usepackage{bbding, wasysym}

\begin{document}

\title{Properties of the Lowest Metallicity Galaxies Over the Redshift Range $z=0.2$ to $z=1$}

\author[0000-0003-4323-0597]{Isaac H. Laseter}
\affiliation{Department of Astronomy, University of Wisconsin-Madison, 475 North Charter Street, Madison, WI 53706, USA}
\author[0000-0002-3306-1606]{Amy J. Barger}
\affiliation{Department of Astronomy, University of Wisconsin-Madison, 475 North Charter Street, Madison, WI 53706, USA}
\affiliation{Institute for Astronomy, University of Hawaii, 2680 Woodlawn Drive, Honolulu, HI 96822, USA}
\affiliation{Department of Physics and Astronomy, University of Hawaii, 2505 Correa Road, Honolulu, HI 96822, USA}
\author[0000-0002-6319-1575]{Lennox L. Cowie}
\affiliation{Institute for Astronomy, University of Hawaii, 2680 Woodlawn Drive, Honolulu, HI 96822, USA}
\author[0000-0003-1282-7454]{Anthony J. Taylor}
\affiliation{Department of Astronomy, University of Wisconsin-Madison, 475 North Charter Street, Madison, WI 53706, USA}

\begin{abstract}

Low-metallicity galaxies may provide key insights into the evolutionary history of galaxies. 
Galaxies with strong emission lines and high equivalent widths 
(rest-frame EW(H$\beta) \gtrsim 30$~\AA) are ideal candidates for the lowest metallicity galaxies to $z \sim 1$.
Using a Keck/DEIMOS spectral database of about 18,000 galaxies between $z=0.2$ and $z=1$, we search for such extreme 
emission-line galaxies with the goal of determining their metallicities.  
Using the robust direct $T_e$ method, we identify 8 new extremely metal-poor galaxies (XMPGs)
with $12 + \log$ O/H $\leq$ $7.65$, including one at $6.949 \pm 0.091$, making it the 
lowest metallicity galaxy reported to date at these redshifts.
We also improve upon the metallicities for two other XMPGs from previous work. 
We investigate the evolution of H$\beta$ using both instantaneous and continuous 
starburst models, finding that XMPGs
are best characterized by continuous starburst models. Finally,
we study the dependence on age of the build-up of metals and the emission-line strength.

\end{abstract}


%***********************
\section{Introduction} 
\label{introduction}
%***********************
Low-redshift ($z\lesssim1$), low-metallicity galaxies could be the best analogs of small,
star-forming galaxies (SFGs) at high redshifts ($z\gtrsim6$).
Such low-redshift galaxies may give insights into the properties of early SFGs,
which may be responsible for reionizing the universe.
\cite{Izotov_2021} found that the properties of low-metallicity, compact SFGs with 
high equivalent widths (EWs) of the H$\beta$ emission line are similar to the properties of high-redshift SFGs.

The lowest metallicity galaxies are often called
extremely metal-poor galaxies (XMPGs).
These are defined as galaxies with metallicities $12 + \log$ O/H $\leq$ $7.65$, which 
is $1/10$th of the solar metallicity (8.65; \citealp{solar_composition}).
A few dozen local XMPGs have been discovered, with the lowest metallicity galaxies being IZw18  
at 7.2 \citep{S&S_1970}, SBS 0335-052W at 7.1 \citep{Izotov_2005}, 
and J0811+4730 at 7.0 \citep{Izotov_2018}. 
%More recent work gives metallicities down to 6.86 \citep{nakajima_2022}.
Recent work by \cite{Senchyna} report metallicities down to 7.25, and \cite{Nakajima_2022} down to 6.86.

For $z\lesssim 1$, XMPGs can be found by targeting extreme emission-line galaxies (EELGs) 
with high rest-frame EWs of H$\beta$ (hereafter, EW$_0$(H$\beta)>30$~\AA). 
To achieve such high EWs, galaxies must have intensive star formation. 
EELGs are relatively common at $z=0.2$--1
 \citep[e.g.,][]{Hoyos_2005,kakazu_2007, Hu_2009, Salzer_2009, Salzer_2020, Ly_2014, Amor_2014, Amor_2015} 
 and contain a significant fraction ($\sim5$\%) of the universal star formation at these redshifts 
 \citep{kakazu_2007, Hu_2009}. The lowest metallicity galaxies from the EELG samples between 
 $z=0.2$--1 are KHC912-29 at $\rm 12 + log(O/H) = 6.97 \pm 0.17$ and KHC912-269 at $7.25 \pm 0.03$ \citep{Hu_2009}.
Low-redshift EELGs have been found with narrowband, grism, and color selections. 
An example of the latter are the Green Pea galaxies found in shallow Sloan Digital Sky Survey (SDSS) 
broadband imaging data \citep{Cardamone_2009}.
A recent analysis of color selections using deep Subaru/Hyper Suprime-Cam broadband imaging data 
can be found in \cite{Rosenwasser_2022}.

Here we use a different approach to the problem, which is more analogous to choosing objects out of the SDSS 
spectroscopic sample but at considerably fainter magnitudes.
We work with an archival sample of galaxies with spectra from Keck/DEIMOS, selecting
those with high EW$_0$(H$\beta$) and redshifts $z=0.2$--1.
Our main goal is to obtain a significant sample of XMPGs with strong [OIII]$\lambda$4363 detection
where we can use the ``direct $T_e$ method" 
\citep[e.g.,][]{Seaton_1975, Pagel_1992MNRAS, Pilyugin_2005ApJ, Yin_2006, Izotov_2006, kakazu_2007, Liang_2007, Hu_2009}.

We organize the paper as follows: In Section~\ref{secspec}, we describe the spectroscopic 
observations. In Section~\ref{secanalysis}, we present our EW and flux measurements,
our electron temperature determinations, our metal abundance measurements, and our 
error analysis. In Section~\ref{secdisc}, we 
give our final XMPG catalog and discuss our results. 
We give a summary in Section \ref{Summary}. 
We use a standard H$_0 = 70$ km~s$^{-1}$Mpc$^{-1}$,
$\Omega_m = 0.3$, $\Omega_{\Lambda} = 0.7$ cosmology 
throughout the paper.

%***********************
\section{Spectroscopic Observations} 
\label{secspec}
%***********************
During the last $\sim20$ years, 
our team obtained spectroscopy of galaxies in a number of well-studied fields 
(e.g., GOODS, COSMOS, SSA22, the North Ecliptic Pole, etc.) for a variety of projects 
\citep[e.g.,][]{Cowie_2004, Cowie_2016, kakazu_2007, Cowie_2008, Barger_2008, Amy_and_Izak_2012, Izak_z_1, Wold_2017, Rosenwasser_2022} 
using the Deep Extragalactic Imaging Multi-Object Spectrograph \citep[DEIMOS;][]{Faber_2003} 
on the Keck~II 10~m telescope.
The observational set-up of the ZD600 line mm$^{-1}$ grating blazed at 7500~\AA\ and a 
$1''$ slit gives a spectral resolution of $\sim 4.5$~\AA\ and a wavelength coverage of 5300~\AA.
However, for an individual spectrum, the specific wavelength range is dependent on the slit position 
with respect to the center of the mask along the dispersion direction. The spectra have an average 
central wavelength of 7200~\AA. 

\begin{figure*}[htb!]
\hskip -0.5cm
\includegraphics[width=7.35in]{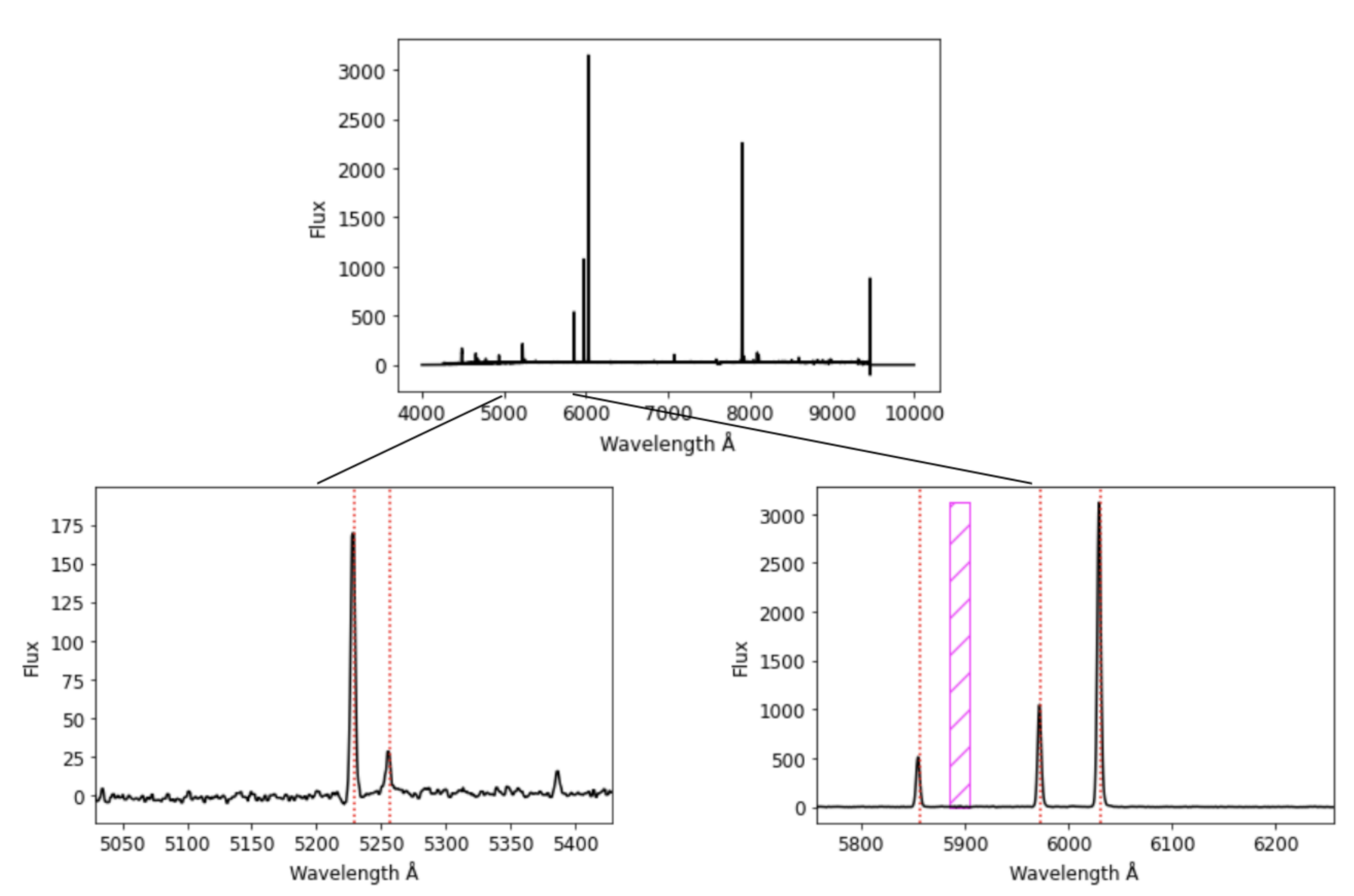} 
\caption{Example DEIMOS spectrum for a galaxy in our sample  
(R.A.(J$2000$)$=$3 32 41.8, Decl.(J$2000$)$=-$ 28 11 25.10,
and $z = 0.2044$). \textit{(Top)} Complete spectrum. \textit{(Bottom left)} Portion of spectrum covering H$\gamma$ and [OIII]$\lambda4363$. \textit{(Bottom right)} Portion of spectrum covering H$\beta$, [OIII]$\lambda4959$, and [OIII]$\lambda5007$. The red dotted lines correspond to
the wavelengths of the lines at the measured redshift.
The pink hatched region marks the location of a sky line. The flux units are arbitrary, as the DEIMOS 
spectra are not photometrically calibrated.}
\label{fig:Deimos_spectra}
\end{figure*}

The observations were not generally taken at parallactic angle, 
since the position angle was determined by the mask orientation. Each $\sim$60~minute exposure 
was broken into three 20~minute subsets, with the objects dithered along the slit by $1\farcs5$ in 
each direction. The spectra were reduced and extracted using the procedures described in \cite{Cowie_1996}. 
In Figure~\ref{fig:Deimos_spectra}, we show an example spectrum to illustrate the quality of the 
DEIMOS spectra obtained.

The overall spectral sample consists of $\sim 52,000$ galaxies with measured redshifts, of which 
$\sim$18,000 lie between $z=0$ and $z=1$. We performed a preliminary  EW$_0$(H$\beta$) 
measurement for all $\sim 18,000$ galaxies to obtain a high EW$_0$(H$\beta$) sample. 
We found 435 galaxies with EW$_0$(H$\beta)>30$~\AA. We next removed any galaxies with
incomplete spectral coverage of [OIII]$\lambda \lambda 5007, 4959$ or
[OII]$\lambda3727$, or where spectral lines are affected by Telluric contamination 
(e.g., 5579~\AA, 5895~\AA, 6302~\AA, 7246~\AA, A-band (7600~\AA --7630~\AA), 
and B-band (6860~\AA --6890~\AA)), which resulted in a final sample of 380 galaxies. 
We hereafter refer to this as ``our sample". 

\begin{figure}[htb!]
\hskip -0.2cm
\includegraphics[width=\columnwidth]{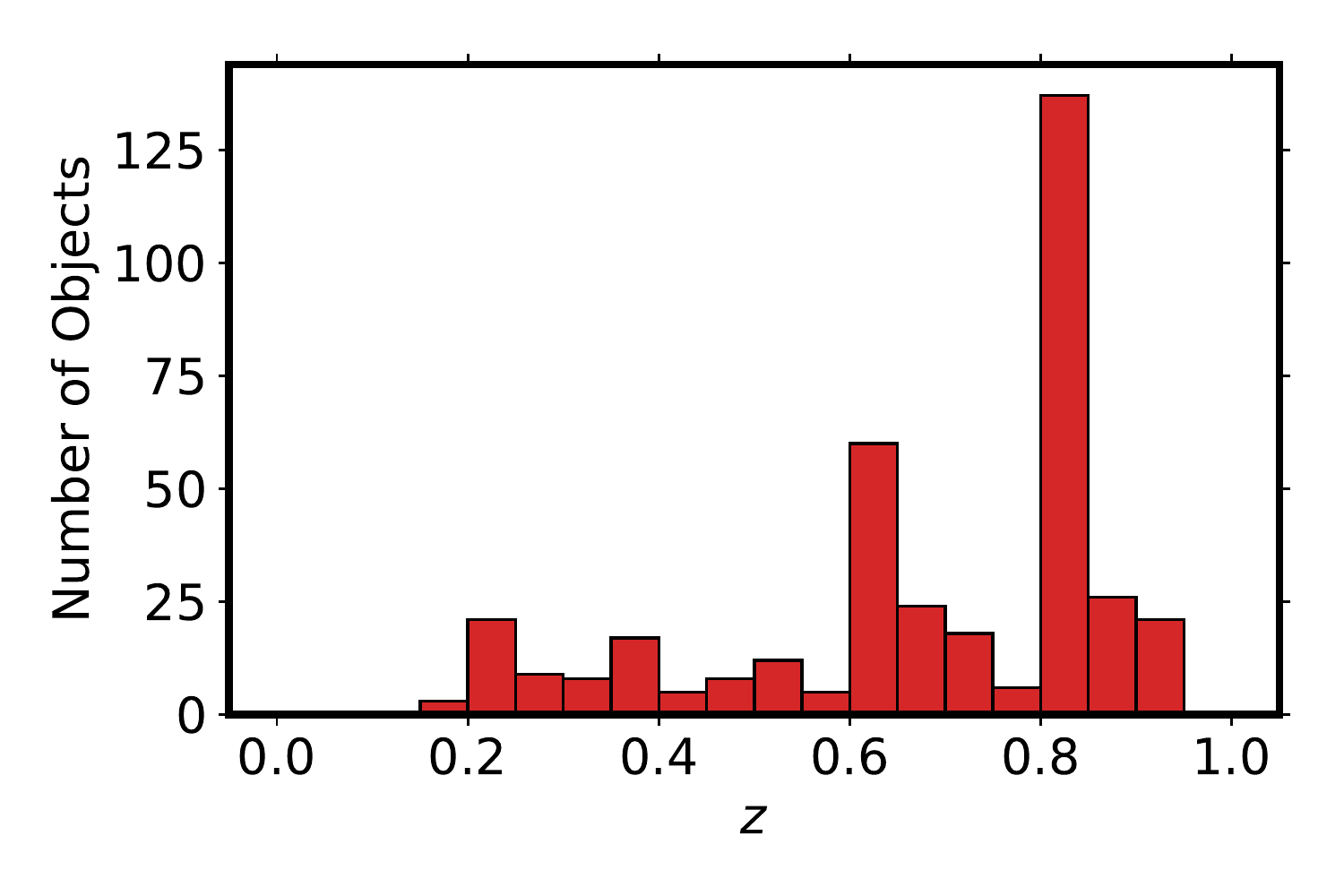} 
\caption{Redshift distribution for our high EW sample (EW$_0$(H$\beta)>30$~\AA) of 380 galaxies. The bin width is $0.05$ in redshift. The large excess present at $z=0.63$ and $z=0.83$ is due to overlap of H$\beta$ and [OIII]$\lambda 5007$ in the narrowband filters NB816 and NB921, respectively.}
\label{fig:Redshift Histogram}
\end{figure}

In Figure \ref{fig:Redshift Histogram}, we present the redshift distribution for our sample. 
There are two peaks that correspond to narrowband filter selection
at $z=0.63$ and $z=0.83$. A substantial fraction of the objects, about 45\%, are in these two peaks.

We note in passing that there are no signs of lensing in the sample. 
We do not see multiple emission lines in the spectra, and the objects are generally 
isolated. Since we are not concerned with luminosities and masses here, lensing would not affect our analysis.

%***********************
% FIGURE 1
%***********************

%***********************
% FIGURE 2
%***********************

%***********************
% FIGURE 3
%***********************
\begin{figure}[h!]
\includegraphics[width=\columnwidth]{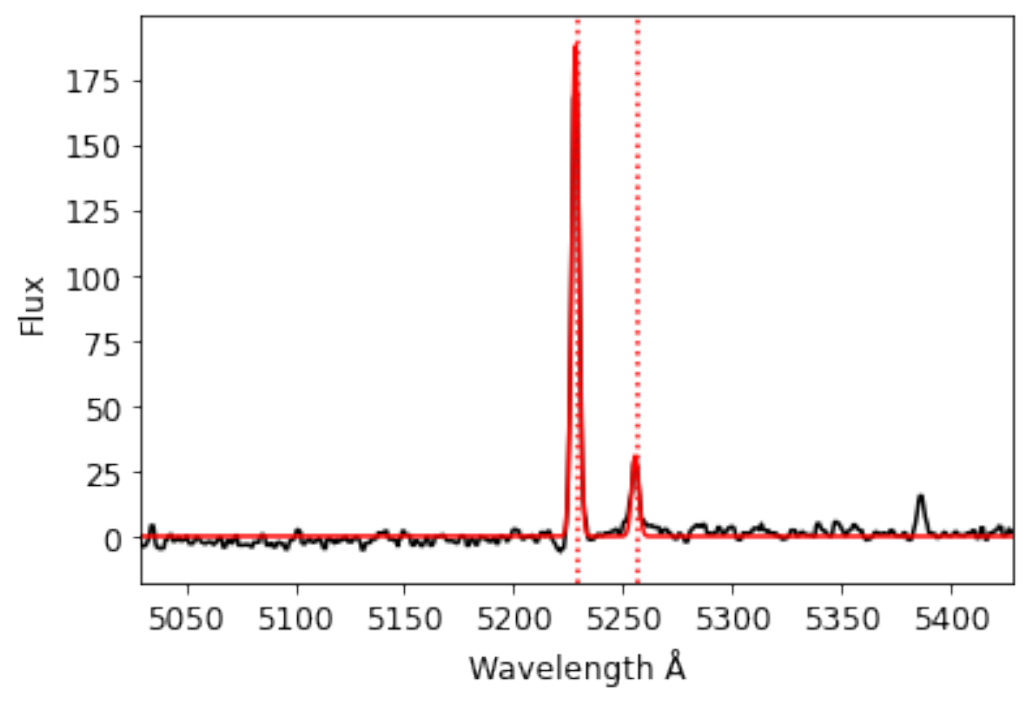} 
\caption{H$\gamma$ and [OIII]$\lambda4363$ emission line fit for the example galaxy in Figure \ref{fig:Deimos_spectra}. The red curve shows the fitted model that the EWs are determined from. 
The red dotted lines show the means of the fitted Gaussians.}
\label{fig:ew example}
\end{figure}

%***********************
% FIGURE 4
%***********************
\begin{figure}
\includegraphics[width=\columnwidth]{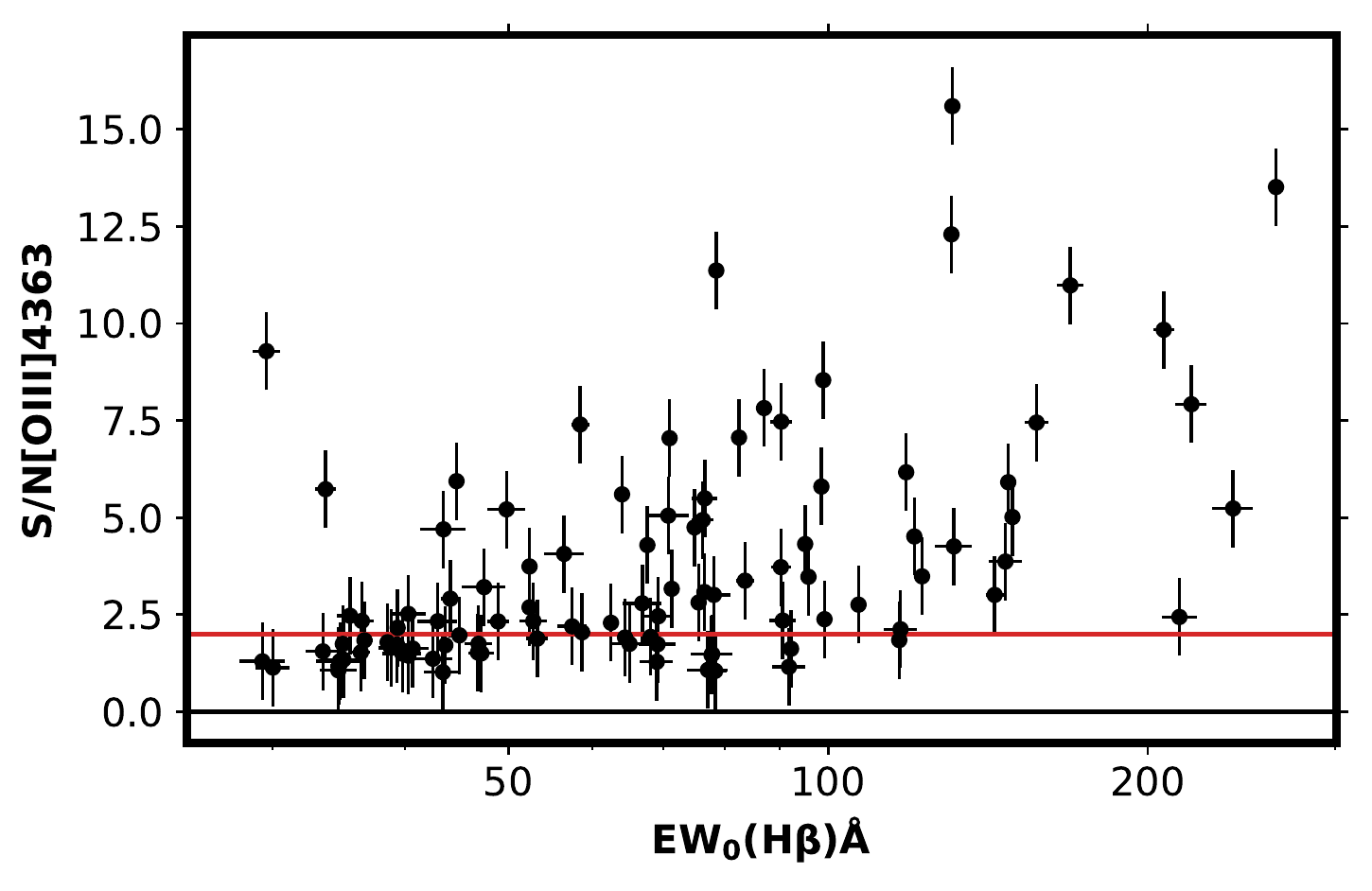} 
\caption{S/N of [OIII]$\lambda4363$ vs. EW$_0$(H$\beta$) for all the sources in our sample where 
H$\beta$ is detected above a S/N of 20. 
The red line shows a S/N of 2 in [OIII]$\lambda4363$.
}
\label{fig:signaltonoise}
\end{figure}

%***********************
% FIGURE 5
%***********************
\begin{figure}
\includegraphics[width=\columnwidth]{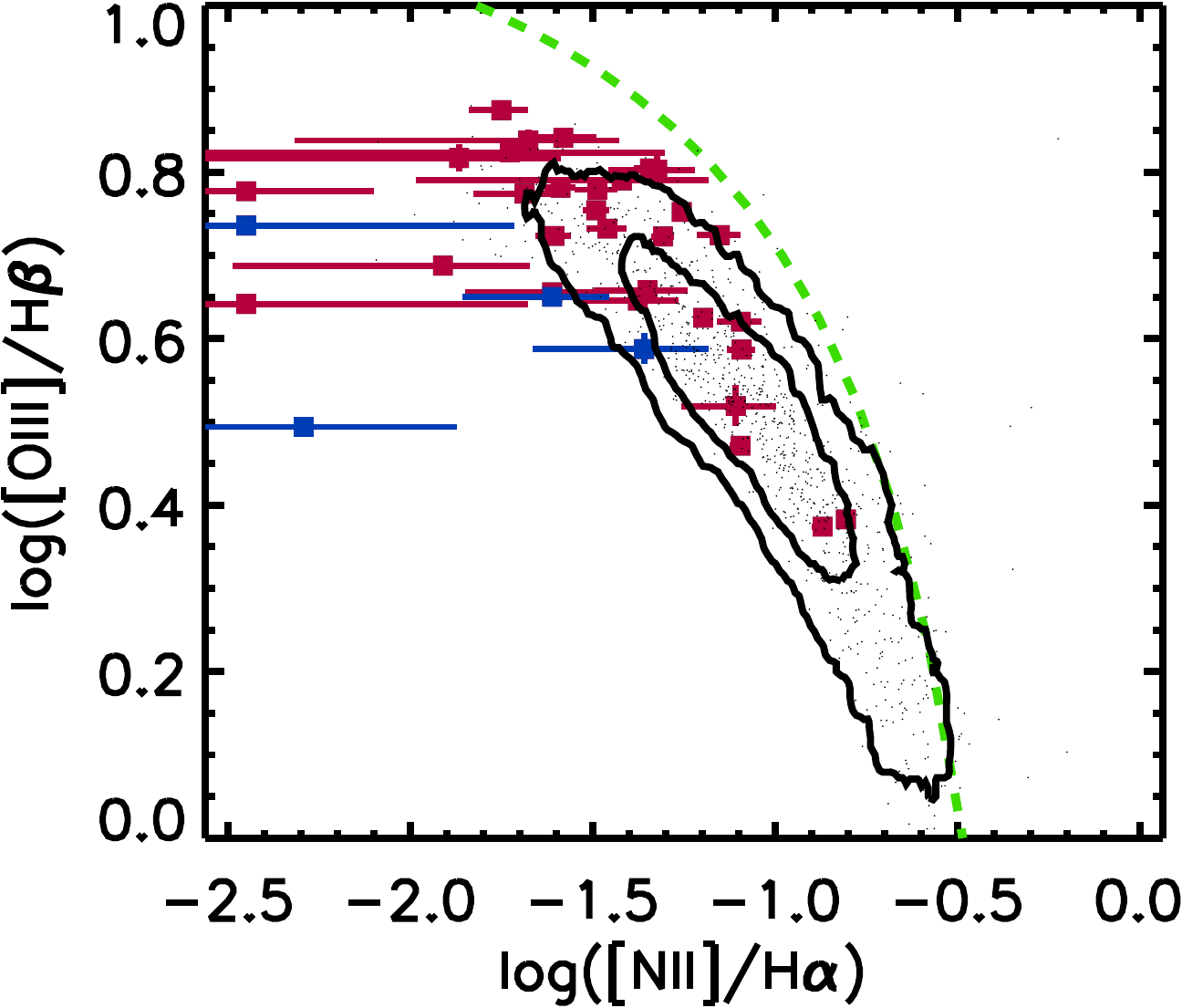} 
\caption{BPT diagram for all the sources in our sample where the H$\alpha $ line is on the 
spectrum and detected above a S/N of 20. The number density contours show the locus 
determined from the SDSS spectral sample. The dots show a subset of the SDSS data points. 
The green dashed curve shows the \cite{Kauff_2003} separation between SFGs and AGNs.
The red and blue points represent galaxies in our sample with 
$\rm 12 + \log (O/H) > 7.65$ and $\rm 12 + \log (O/H) \leq 7.65$, respectively.
}
\label{fig:bpt}
\end{figure}

%***********************
% FIGURE 6
%***********************
\begin{figure}
\includegraphics[width=\columnwidth]{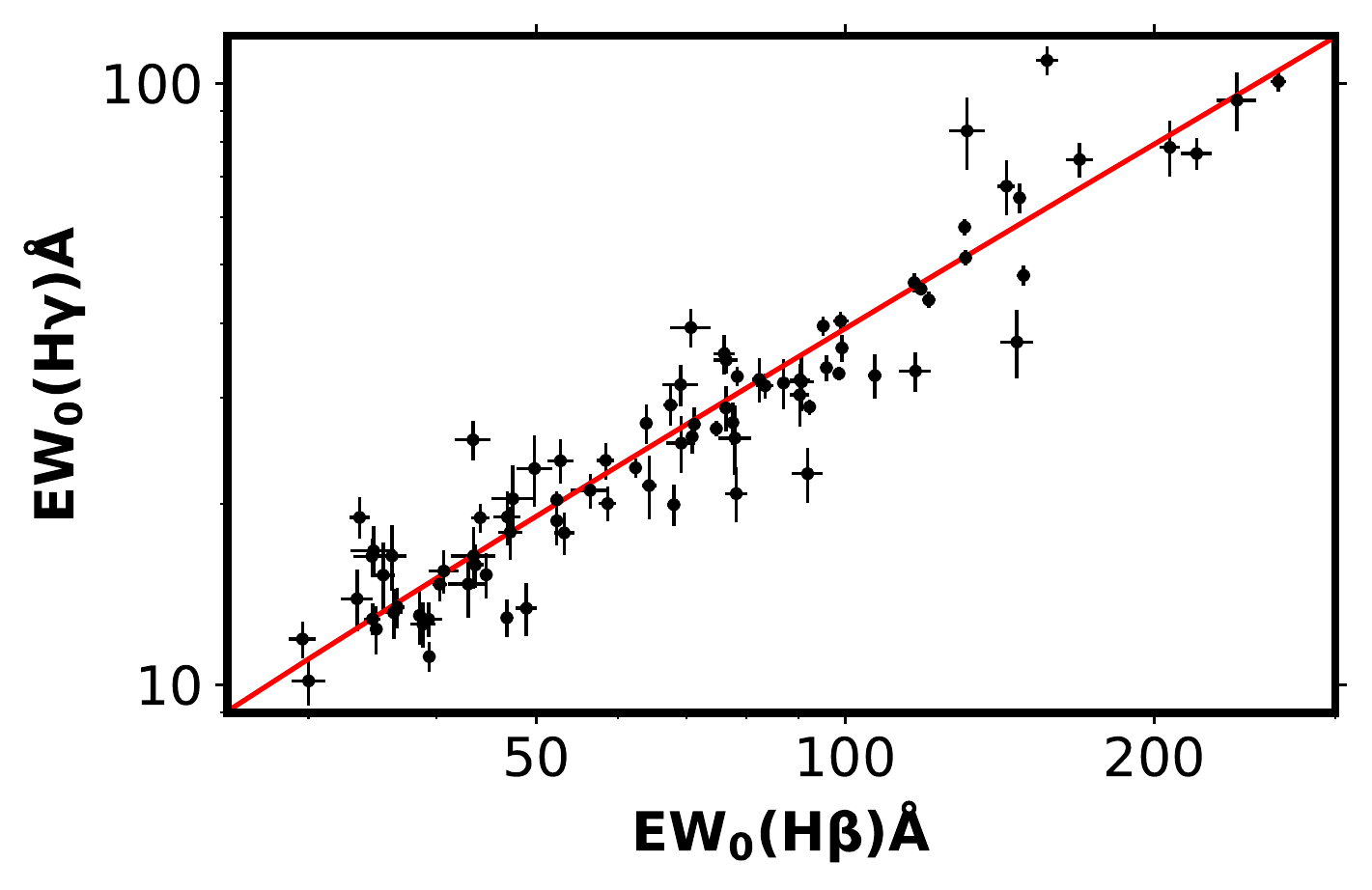} 
\caption{EW$_0$(H$\gamma$)
vs. EW$_0$(H$\beta$) for all the sources in our sample where H$\beta$ is detected above a S/N of 20.
The Balmer decrement is shown by the red line, assuming the underlying continuum is flat in $f_\nu$.
}
\label{fig:ewhb_balmer}
\end{figure}

%***********************
\section{Analysis} 
\label{secanalysis}
%***********************

%***********************
\subsection{EW Measurements} 
\label{secEW}
%***********************
To determine the metallicities of our sample, we measure the EWs of the emission lines of interest 
in each spectrum: [OII]$\lambda3727$, [OIII]$\lambda4363$, [OIII]$\lambda5007$, and [NII]$\lambda6583$,
as well as H$\gamma$, H$\beta$, and H$\alpha$.
We use EWs instead of emission line fluxes to avoid introducing biases from spectral flux calibrations. 
A challenge with determining EWs is measuring the continua of the spectra. We carefully inspected the 
spectra of all the sources with very weak continua to make sure that the continua are plausible.
However, since we are primarily concerned with local line ratios, a continuum measurement
is not critical to the metallicity determination.
To measure the EWs, we simultaneously fit Gaussian functions to neighboring emission lines of interest, 
and we divide the integrated area under each fitted Gaussian by the median continuum level 
near each line. To convert these observed-frame EWs to rest-frame EWs, we divide by $(1+z)$.

When fitting H$\beta$ and the [OIII]$\lambda\lambda4959,5007$ doublet, we enforce a 3:1 internal 
ratio for the doublet and require that all three lines have the same line width. We fit a single redshift 
and assume that the FWHM is the same for all three lines. We fit the lines simultaneously using four 
free parameters ($z$, FWHM, and the amplitudes of the [OIII]$\lambda5007$ and H$\beta$ lines).
We use the same procedure to fit simultaneously the H$\alpha$ line and the [NII]$\lambda\lambda6548,6583$ 
doublet. We also use similar procedures to fit two Gaussian functions, with a single FWHM and redshift, 
to the H$\gamma$ and [OIII]$\lambda4363$ lines, as well as to the [OII]$\lambda\lambda3729,3727$ doublet. 
We do not account for stellar absorption in the Balmer lines. Given the strength of the emission lines in
EELGs, the effects of stellar absorption are extremely small.

By fitting multiple Gaussians simultaneously rather than individual lines independently, we increase the fidelity of the fits to weaker emission features. For example, [OIII]$\lambda4363$ is a relatively weak line compared to the neighboring H$\gamma$ line. However, by simultaneously fitting H$\gamma$ and [OIII]$\lambda4363$, we can infer the line center and line width from the stronger H$\gamma$ line, thereby improving the fit to the weaker [OIII]$\lambda4363$ line. In Figure~\ref{fig:ew example}, we show the emission line fit for the
example galaxy shown in Figure~\ref{fig:Deimos_spectra}. 

Selecting EELGs increases the likelihood of getting [OIII]$\lambda4363$ detections. 
In Figure~\ref{fig:signaltonoise}, we show all of the sources in our sample where H$\beta$ is 
detected above a S/N of 20. 
For EW$_0$(H$\beta$)$>50$~\AA, 46 of the 70 sources are detected above a $2\sigma$ level in
the [OIII]$\lambda4363$ line.
For EW$_0$(H$\beta$)$>100$~\AA, only 4 of the 22 sources do not have $>2\sigma$ 
detections in [OIII]$\lambda4363$, and most sources have a very strong detection.

In Figure~\ref{fig:bpt}, we show a Baldwin, Phillips, \& Terlevich (BPT) diagram \citep{Baldwin_1981} 
for the lower redshift sources in our sample that have H$\alpha$ measurements. We require S/N $>20$ 
in H$\alpha$ to choose sources where we can make a robust measurement of [NII]/H$\alpha$. 
We compare with the locus determined by the SDSS spectral sample. The figure demonstrates that 
our EW measurements are robust and that we have a spectral sample of SFGs that follow the BPT track. 
Although Figure~\ref{fig:bpt} is limited to lower redshift sources due to the spectral coverage of [NII] and 
H$\alpha$, our higher redshift sources also do not show active galactic nucleus (AGN) signatures, 
such as $\rm [NeV]\lambda 3426$. This suggests relatively little AGN contamination in the higher redshift 
part of the sample.

%***********************
\subsection{Flux Determinations} 
\label{secflux}
%***********************
To convert the [OIII]$\lambda5007$ to [OIII]$\lambda4363$ EW ratio to a flux ratio, we use the relation
\begin{equation} 
\label{5007/4363}
    \frac{f{\rm (\lambda5007})}{f{\rm (\lambda4363)}} = \frac{\rm EW( \lambda5007)}{\rm EW(H\beta)}\times\frac{\rm EW(H\gamma)}{\rm EW(\lambda4363)}\times\frac{f(H\beta)}{f(H\gamma)} \,.
\end{equation}
We assume the extinction in the sources is low and fix the flux ratio of H$\gamma$ to H$\beta$ 
at the Balmer decrement value of 0.47 (e.g., \citealt{Osterbrock_1989}).
We also assume the continuum is flat between adjacent lines.

We may also compute flux ratios by assuming a shape for the underlying continuum. 
In Figure~\ref{fig:ewhb_balmer}, we plot EW$_0$(H$\gamma$)
versus EW$_0$(H$\beta$) for sources with H$\beta$ S/N $>20$.
There is a near linear relation. Assuming the underlying continuum is flat in $f_\nu$
gives a median flux ratio for H$\gamma$ to H$\beta$ of 0.465, consistent with the Balmer 
decrement, which is shown by the red line.
We use this method to compute the flux ratio of the combined [OII]$\lambda\lambda3729,3727$ 
doublet to the H$\beta$ line. We consider this method more uncertain than the Balmer decrement
method, which we use for the other line ratios.

%***********************
\subsection{Electron Temperature and Oxygen Abundance Determinations} 
\label{secTe}
%***********************
In the $T_e$ method, the metallicities of galaxies are deduced from the flux ratio of the [OIII]$\lambda\lambda4959,5007$ doublet to the [OIII]$\lambda4363$ thermal line.
We follow the prescription of  \citet{Izotov_2006} and use their Equations~1 and 2 to 
determine the electron temperatures.

Using the $T_e$ method also allows ion abundances for 
galaxies to be derived directly from the strength of the emission lines,
specifically O/H. \cite{Izotov_2006} empirically fit relations to an electron temperature range for common galactic emission lines. This results in a large number of ionic abundance relations, including the desired O$^+$ and O$^{2+}$ relations for this work (their Equations~3 and 5). To obtain the total abundance of Oxygen for each galaxy, these two equations must be added together. Note that there is typically 
little change in the total metallicity when adding O$^+$ to O$^{2+}$, as O$^{2+}$ is the strongly
dominant ionization state of Oxygen in these galaxies.
This means any uncertainties in our conversion of the EW ratio to flux ratio for 
[OII]$\lambda3727$/H$\beta$ are less important.

%***********************
\subsection{Error Analysis} \label{error}
%***********************
We first calculated the errors on the multi-Gaussian continuum fits. For each group
 of simultaneously fit emission lines, we shifted the positions of the fitted lines along the spectra in regular intervals, refitting the emission lines at each interval and calculating EWs. We then took the standard deviations of these EW measures as the $1\sigma$ error bounds on the EWs of the lines. 
 We used this method in order to properly sample the systemic, non-uniform noise in the spectra resulting from sky subtraction procedures. 

To propagate the measurement errors 
we used a Monte Carlo technique, as analytical propagation would be needlessly complex. 
In this technique, we 
evaluated the Oxygen abundance $10,000$ times, using values drawn randomly from normal distributions for the EWs of [OIII]$\lambda\lambda\lambda$5007,4959,4363 H$\beta$, H$\gamma$, and [OII]$\lambda3727$, centered at the measured values, and with standard deviations corresponding to the $1\sigma$ errors on each value. We used the standard deviation of the distribution of total abundance for each galaxy as the $1\sigma$ error on the abundance of that galaxy. A general overview of the above error propagation procedure can be found in \cite{andrae2010error}.

%***********************
\section{Discussion} 
\label{secdisc}
\begin{figure*}[htb!]
\centering
\includegraphics[width=\columnwidth]{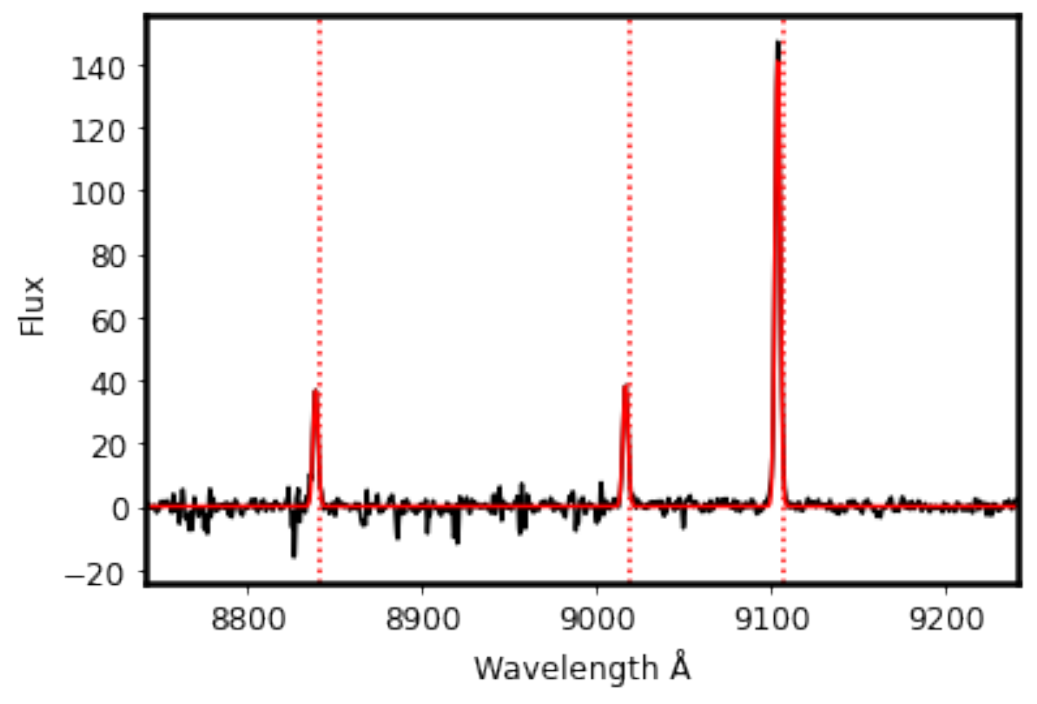} 
\includegraphics[width=\columnwidth]{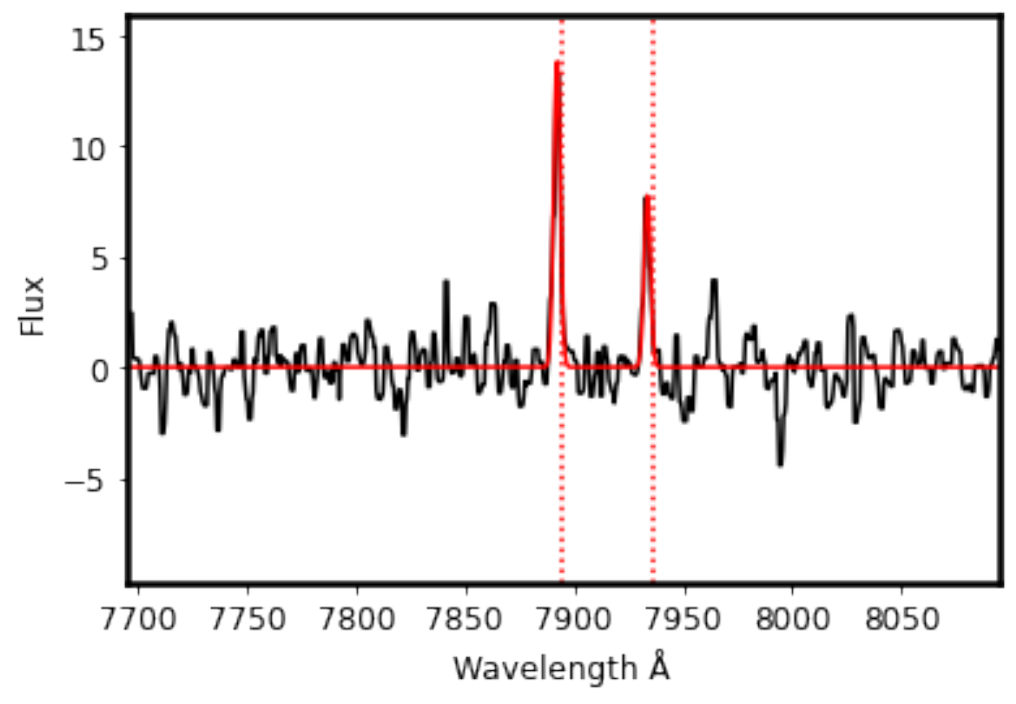} 
\includegraphics[width=\columnwidth]{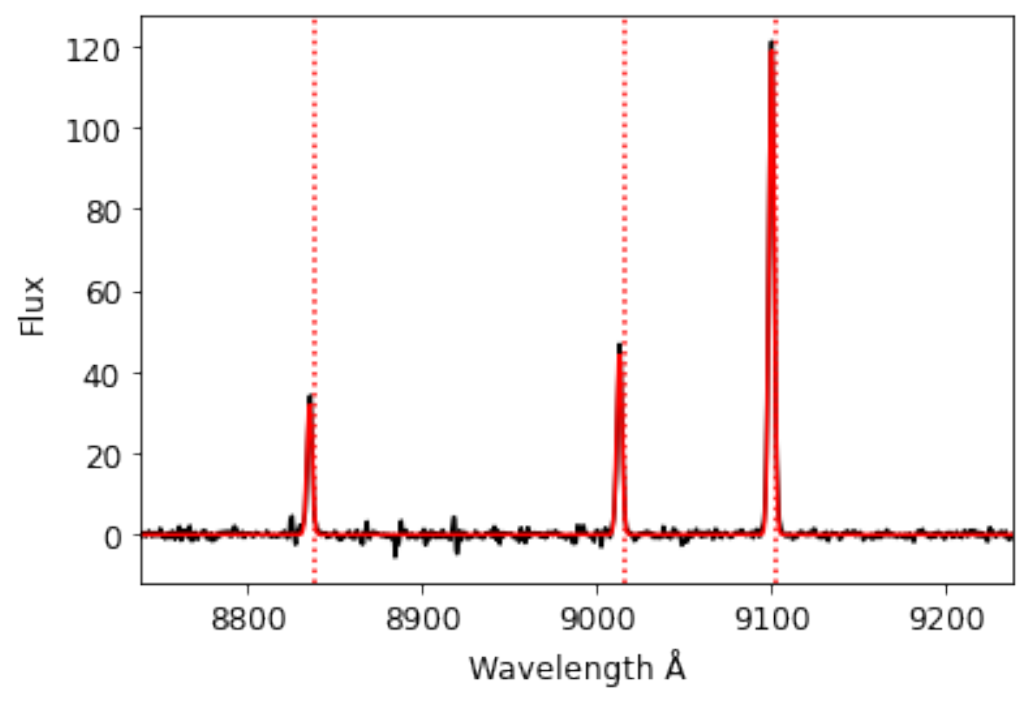} 
\includegraphics[width=\columnwidth]{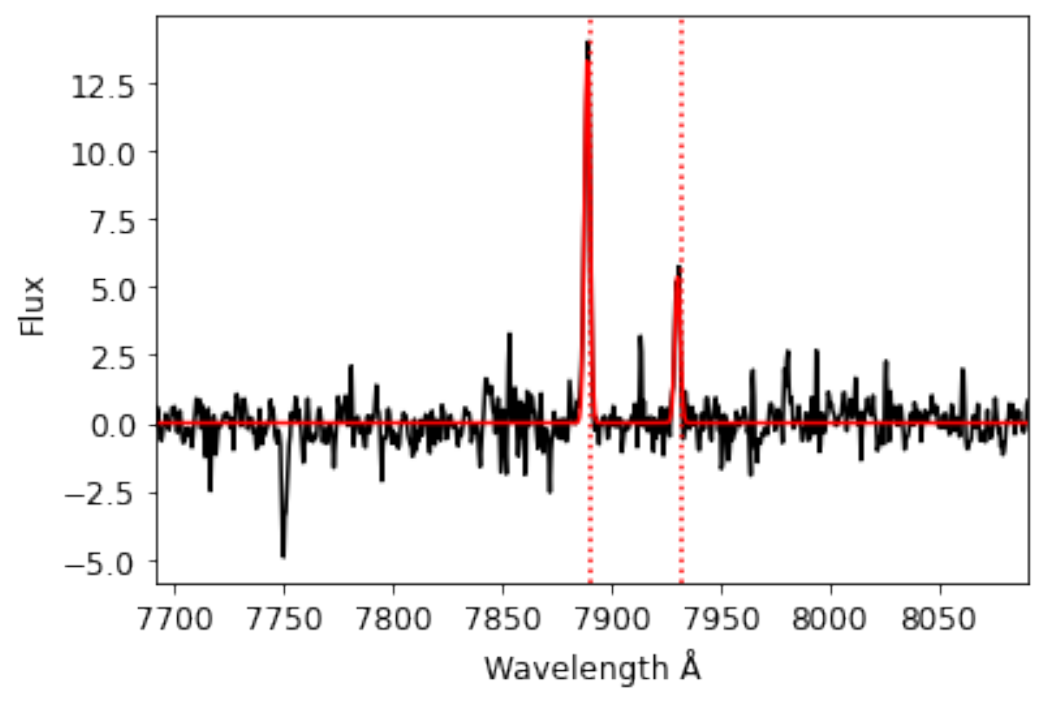}
\caption{\textit{Left}: Emission lines H$\beta$, [OIII]$\lambda 4959$, and [OIII]$\lambda 5007$ 
in the (top) lowest metallicity galaxy with $12 +$log O/H $=6.949 \pm 0.091$ and
(bottom) second lowest metallicity galaxy with $12 +$log O/H $=7.208 \pm 0.061$,
along with their triple Gaussian fits. 
\textit{Right}: Emission lines H$\gamma$ and [OIII]$\lambda 4363$ for the same two galaxies,
along with their double Gaussian fits. A 5-point median smoothing 
has been applied. 
}
\label{fig:lowest metallicity}
\end{figure*}

%***********************
Following the procedures in Section~\ref{secanalysis},
we determined the $12 +$log O/H abundance for each galaxy in our sample
with S/N $>20$ in the H$\beta$ line and
detected above a $1\sigma$ threshold in [OIII]$\lambda 4363$. 
In Table~\ref{table:only_table}, we summarize  the 10 XMPGs found in this way.
The two lowest metallicities are $12 +$log O/H $=6.949 \pm 0.091$ and 
$7.208 \pm 0.061$ at EW$_0$(H$\beta$) of $169 \pm 5$~\AA{} and $264 \pm 4.69 $~\AA,
respectively. In Figure~\ref{fig:lowest metallicity}, we show the triple Gaussian fits to 
H$\beta$, [OIII]$\lambda 4959$, 
and [OIII]$\lambda 5007$ and the double Gaussian fits to H$\gamma$ and [OIII]$\lambda 4363$ 
for these two galaxies.
We show the spectra of the remaining XMPGs in the Appendix, along with $z$-band images of
the four XMPGs in the SSA22 field using newly reduced Subaru Hyper Suprime-Cam imaging
(B. Radzom et al.\ 2022, in preparation).

A majority of our objects have been identified in previous work. However, our present spectroscopic observations 
are significantly improved over those observations and allow for the 
detection of the [OIII]$\rm \lambda 4363$~\AA\ line. Thus, our metallicity measurements for most of the objects are new. 
Two of our objects had previous metallicity measurements determined by \cite{kakazu_2007}. 
In one case (2$^h$ 40$^m$ 35.56$^{s}$, $-1^\circ$ $35^{'}$ $37.1^{''}$),
the present 5~hr exposure is comparable to the 4~hr exposure used in \cite{kakazu_2007}.
In the other case (22$^h$ 19$^m$ 06.30$^{s}$, $+0^\circ$ $47^{'}$ $21.6^{''}$),
the present 7~hr exposure is considerably longer than the 1~hr exposure used in \cite{kakazu_2007},
and the errors are correspondingly lower.
We include in Table \ref{table:only_table}, if available, the original source identification and the original metallicity determination.

%***********************
% TABLE 1
%***********************
\begin{longrotatetable}
\begin{deluxetable}{lrccccccccc}
\tablecaption{All 10 XMPGs Discovered from the Final Spectral Sample of 380 Galaxies\label{table:only_table}}
\scriptsize
\tablehead{
R.A. &  Decl. &  $z_{spec}$ & \multicolumn{5}{c}{EW$_0${(\AA)}}  & $12 +$ log(O/H) & Field & Exp \\
\multicolumn{2}{c}{(J2000.0)} &  & H$\gamma$ & [OIII]$\lambda4363$ & H$\beta$ & [OIII]$\lambda4959$ & [OIII]$\lambda5007$ & & & (hr)  \\
(1) & (2) & (3) & (4) & (5) & (6) & (7) & (8) & (9) & (10) & (11)}
\startdata
\hline\hline
22 17 46.38$^{\dagger}$ & +0 18 13.0 &  0.8183 &  $74.743\pm4.939$ &      $41.609\pm3.789$ &   $169.075\pm5.004$ &     $217.107\pm10.03$ &     $651.32\pm10.023$ &   $6.949\pm0.092$ & SSA22 & 19  \\
22 19 06.30$^{\dagger \dagger}$ & +0 47 21.6 &  0.8175 &  $100.766\pm3.98$ &      $40.377\pm2.987$ &   $264.084\pm4.676$ &     $328.585\pm6.434$ &     $985.756\pm7.554$ &   $7.208\pm0.061$ & SSA22 & 7 \\
12 37 31.51$^{\ast}$ & +62 10 6.0 &  0.1724 &  $16.402\pm1.933$ &       $8.704\pm1.853$ &    $43.443\pm2.155$ &       $78.84\pm1.691$ &     $236.521\pm1.618$ &    $7.326\pm0.18$ & HDFN & 1 \\
02 41 31.81$^{\dagger}$ & -1 33 14.8 &  0.3930 &  $67.517\pm7.056$ &      $13.113\pm4.359$ &     $143.494\pm2.8$ &       $149.177\pm2.6$ &     $447.532\pm4.029$ &   $7.333\pm0.263$ & XMM-LSS & 17 \\
22 17 45.40$^{\dagger}$ & +0 28 18.4 &  0.6175 &   $20.41\pm2.757$ &        $8.17\pm2.545$ &    $47.439\pm2.255$ &      $72.363\pm4.052$ &      $217.09\pm3.099$ &   $7.338\pm0.256$ & SSA22 & 6 \\
02 40 35.56$^{\dagger \dagger}$ & -1 35 37.1 &  0.8206 &  $22.911\pm3.094$ &      $12.276\pm2.355$ &    $49.809\pm2.001$ &     $102.334\pm1.795$ &     $307.002\pm1.967$ &   $7.445\pm0.167$ & XMM-LSS & 5 \\
03 34 09.78$^{\wr}$ & -28 1 28.0 &  0.8540 &   $93.78\pm10.37$ &      $38.336\pm7.321$ &  $240.602\pm10.555$ &    $393.776\pm12.252$ &   $1181.328\pm14.547$ &   $7.477\pm0.157$ & CDFS & 5 \\
10 44 51.69 & +59 4 22.5 &  0.2295 &  $20.815\pm2.178$ &        $5.28\pm5.025$ &    $78.292\pm1.951$ &      $116.64\pm5.454$ &      $349.92\pm2.273$ &   $7.554\pm0.551$ & Lockman & 1 \\
22 17 27.33$^{\dagger}$ & +0 11 50.5 &  0.3954 &  $12.849\pm3.351$ &       $3.086\pm1.567$ &     $44.97\pm1.518$ &      $70.539\pm2.211$ &     $211.618\pm1.789$ &    $7.61\pm0.455$ & SSA22 &10  \\
10 33 03.85 & +57 58 13.6 &  0.3370 &  $12.965\pm3.511$ &       $2.403\pm1.762$ &    $42.458\pm1.812$ &      $54.751\pm1.233$ &     $164.252\pm1.314$ &   $7.647\pm0.547$ & Lockman & 2
\enddata
\tablecomments{Right ascension is given in hours, minutes, and seconds, and declination is given in degrees, arcminutes, and arcseconds. 
$^{\dagger}$Identified in \cite{kakazu_2007}.
$^{\dagger \dagger}$Objects with direct $\rm T_e$ metallicity determinations in \cite{kakazu_2007}. 
The respective metallicities are $12 +$ log(O/H) $= 7.43 \substack{+0.18 \\ -0.15}$ and $12 +$ log(O/H) $= 7.43 \substack{+0.22 \\ -0.17}$.
$^{\ast}$Identified in \cite{Ashby_2015} 
$^{\wr}$Identified in \cite{Izak_z_1}.}
\end{deluxetable}
\end{longrotatetable}

%***********************
% FIGURE 7
%***********************

\begin{figure*}[htb!]
\centering
\includegraphics[width=\textwidth]{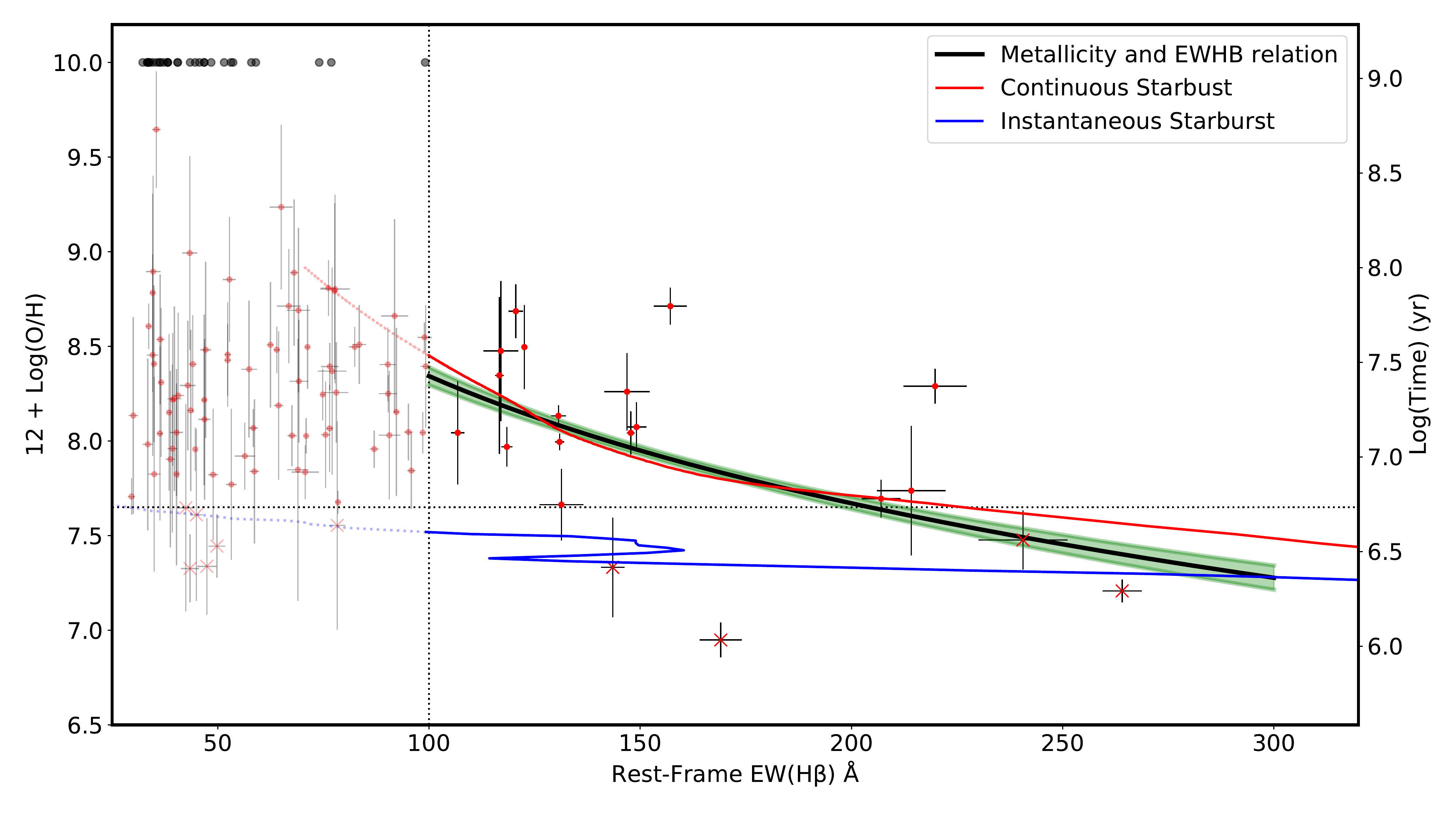}
\caption{
EW$_0$(H$\beta$) vs. metallicity for our sample with S/N cuts of H$\rm \beta \geq 20 $ and [OIII]$\lambda4363$ $\geq 1$. Data that met the S/N cut of H$\beta \geq 20$ but not [OIII] $\lambda4363$ $\geq 1 $ are plotted at a nominal value of $10$. Prior EW$_0$(H$\rm \beta$) vs. metallicity plots suffered from a lack of galaxies populating the region where EW$_0$(H$\beta$) exceeded $\rm 100~\AA$, and thus the overall shape of the relation was under-constrained. Here we see that as the rest-frame EW increases, the metallicity decreases. 
The black vertical dotted line shows the low metallicity threshold of $12+$log (O/H)$=7.65$, which is $1/10$th of the solar metallicity. 
The black solid curve represents a linear fit of $12 + \rm{log(O/H)}  =  12.810 \pm 0.101 -2.233 \pm 0.061  \times \log \rm EW_0(H\beta)~\AA$. We show the $\rm 1\sigma$ distribution of the Monte Carlo simulation fits in green shading. 
The metallicity abundance error and EW$_0$(H$\beta$) error were determined from the procedures presented in Section \ref{error}.
Also shown are the continuous (red curve) and instantaneous (blue curve)
starburst models from \texttt{Starburst99}. 
We plot the data points at EW$_0$(H$\beta)<100$~\AA\ with fainter symbols 
and change each starburst model curve to dotted. We plot the XMPG sample as a different symbol to differentiate between our high metallicity and low metallicity sample.
There is agreement between the metallicity and EW$_0$(H$\beta$) relation 
and the continuous starburst model at high EWs, which underlines the 
dependence of metallicity and EW$_0$(H$\beta$) on age.}
\label{fig:final_plot}
\end{figure*}

In Figure~\ref{fig:final_plot} we show $12 +$log O/H abundance versus EW$_0$(H$\beta$). Those
galaxies not detected above $1\sigma$ in [OIII]$\lambda 4363$ are shown at a nominal 
(high) value of 10. 
In examining the figure, a relation can be seen between $12 +$log O/H 
and EW$_0$(H$\beta$). Specifically, as EW$_0$(H$\beta$) increases, the metallicity decreases
with a median metallicity of 8.39 for sources with EW$_0=50-100$~\AA{} and 8.07 for sources with
EW$_0>100$~\AA{}. 
For the EW$_0\ge100$~\AA{} region, a Pearson's correlation test gives a linear correlation factor (r-value) of $-0.51$ 
and a probability value of $0.02$ for an uncorrelated system to produce the same r-value.

A simple linear fit between the O abundances and $\log$ EW$_0$(H$\beta$) gives the relation
\begin{multline} 
\label{metallicity_ew}
12 + \rm{log(O/H)}  =  12.810 \pm 0.101\\ -2.233
\pm 0.061  \times \log \rm EW_0(H\beta)~\AA
\end{multline}
for galaxies with EW$_0\geq100$~\AA{} (black curve). We determined the errors on the fit by performing a Monte Carlo simulation based on the uncertainties in the EW$_0$($H\beta$) and metallicity of the dataset. We show the $1\sigma$ distribution of the resulting set of fits in green shading in Figure \ref{fig:final_plot}.
The minimum EW$_0$(H$\beta$) requirement ensures the fitted relation 
accurately represents the higher EW$_0$(H$\beta$) lower metallicity region, which has been under-constrained 
in past XMPG surveys. This relation
underlines the effectiveness of searching for XMPGs in samples of high-EW emission-line galaxies.
To highlight the high-EW sample we focus on subsequently, we plot the data points at EW$_0$(H$\beta) < 100$~\AA\ with fainter symbols, and we plot a vertical line at EW$_0$(H$\beta) = 100$~\AA.

%***********************
% FIGURE 8
%***********************

We are interested in the relationships between EW$_0$(H$\beta$), metallicity, 
and galaxy age (t) at EW$_0$(H$\beta)\geq100$~\AA; specifically, the 
EW$_0$(H$\beta$) evolution and the changes in 
metallicity as a function of galaxy age.
By galaxy age, we mean the time since the onset of the currently dominant star formation episode. This
does not preclude there being older underlying populations in the galaxy. 
To determine these relationships, we must first assume a star formation model. 

\begin{figure}
\centering
\includegraphics[scale = 0.5]{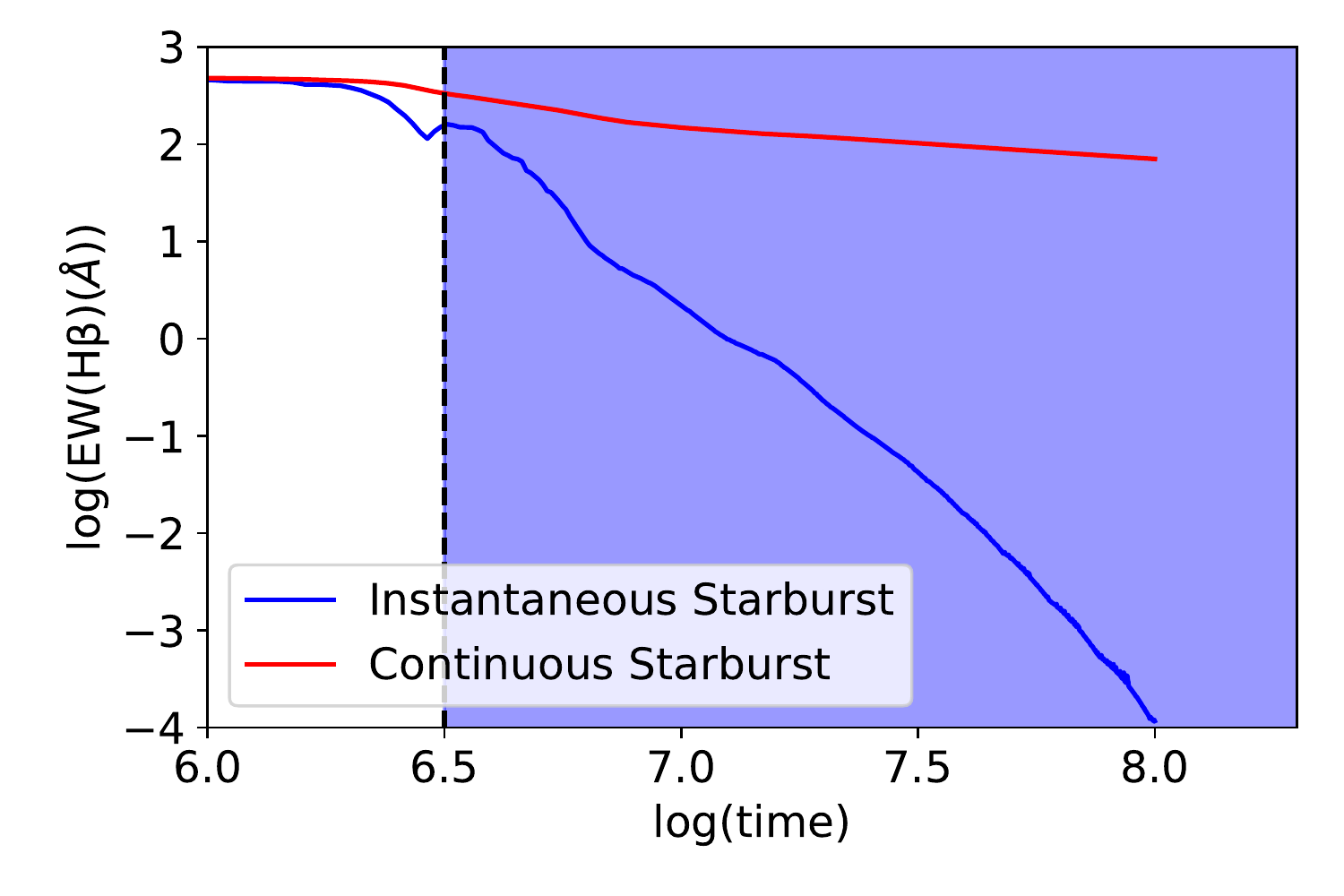} 
\caption{EW(H$\beta$) vs. time for the instantaneous (blue curve) and continuous (red curve)
starburst models from \texttt{Starburst99}. The shaded region represents the time 
region of each model where we fit the models.
\label{fig:starburst}
}
\end{figure}

We constructed instantaneous and continuous starburst models using the program 
\texttt{Starburst99} \citep{Starburst_1999_1, Starburst_1999_2, Starburst_1999_3, Starburst_1999_4}.
We left the initial parameters for each \texttt{Starburst99} model 
unchanged. We show the models in Figure \ref{fig:starburst}.

%***********************
% FIGURE 9
%***********************

In Figure \ref{fig:final_plot}, we see that the instantaneous starburst model, where 
the EW(H$\beta$) drops very rapidly with time, is a poor fit to our data at 
EW$_0$(H$\beta$)$\geq100$~\AA; thus, we focus  our attention on the continuous starburst model.
Specifically, we fit a power law to the continuous starburst model for $\rm t > 10^7$ yr, which
gives the relation
\begin{equation} \label{age_ew}
    \rm \log t (\rm yr) = 12.956 -2.700 \times \log \rm EW(H\beta) \,.
\end{equation}
In addition to our assumption that our galaxies are undergoing continuous starburst,
we assume that (1) the oxygen abundance increases linearly with  time, and (2) the hydrogen abundance
 remains constant throughout time. With these assumptions, we obtain the following relation 
 between metallicity and age:
\begin{equation} \label{age_metallicity}
    12 + \log \rm O/H = \delta + \log t (yr) \,,
\end{equation}
or, combining with Equation~\ref{age_ew},
\begin{equation} \label{eqn_final}
    12 + \log \rm O/H = \delta +  12.956 -2.700 \times \log \rm EW_0(H\beta) \,,
\end{equation}
where $\delta$ is a single fitted parameter, which is a measure of the yield. 

We overplot Equation~\ref{eqn_final} on the distribution of galaxy metallicities
for $\delta=$0.93, which we chose to match the EW$_0$(H$\beta)\ge100$~\AA\ points (red curve). 
The blue curve is the instantaneous starburst model with the $\delta$ offset set to an arbitrary value.
This curve is too flat to provide a fit to the data at EW$_0$(H$\beta)\geq100$~\AA.

The continuous starburst model shows reasonable agreement with the 
EW$\rm _0$(H$\beta) \geq 100~$\AA\ data, but 
it over-predicts the EW$\rm _0$(H$\beta) < 100$~\AA\ data,
suggesting that the effective yield or the specific 
star formation rate is dropping with time.
However, Figure \ref{fig:final_plot} supports the main point, which is that there is a 
clear relation between EW$_0$(H$\beta$), 
age, and metallicity for EW$\rm _0$(H$\beta) \geq 100~$\AA
and that young XMPGs are undergoing star formation rates that are
closer to continuous rather than instantaneous.
Thus, for a galaxy with a measured EW$_0$(H$\beta$)$\geq100$~\AA, we can estimate the galaxy's 
age and the galaxy's metal abundance using Equations~\ref{age_ew} and \ref{eqn_final}.

The models we test here do not exhaust the full range of possible parameter space of \texttt{Starburst99}, 
and thus do not cover the full scope of star formation histories. Nonetheless, the continuous starburst
model fits our high-EW sample well.
The relation between EW$_0$(H$\beta$) and metallicity is present due to the underlying 
relation of the two with the age of the galaxy.

In summary, we show that for EW$_0$(H$\beta$)$\geq100$~\AA,
EW$_0$(H$\beta$) is a good  proxy for galaxy age and metallicity, and 
XMPGs are best modeled by continuous starburst models.

%***********************
\section{Summary} \label{Summary}
%***********************
In this study, we discovered 8 new galaxies below the XMPG threshold of $12 +\log$ O/H $=7.65$ 
and improved upon metallicity measurements from \cite{kakazu_2007} for two more.
Our lowest metallicity galaxy 
has $12 +$log O/H $=6.949 \pm 0.091$. We compared metallicity and EW$_0$(H$\beta$) for 
our spectral sample and found that at EW$_0$(H$\beta$)$\geq100$~\AA, there is
a clear relation between the two, which we interpret as being a
consequence of a near continuous star formation rate in the galaxy.
For these sources, EW$_0$(H$\beta$) is an adequate proxy for galaxy age and metallicity.

With the spectroscopic sample sizes continually increasing, we expect to find even lower 
metallicity galaxies, which will help determine if there is a minimum galaxy metallicity in a 
given redshift range.

%***********************
\section*{Acknowledgements}
%***********************
We thank the anonymous referee for very constructive comments that improved
the manuscript.
We gratefully acknowledge support for this research from the Diermeier Family Foundation Astronomy Fellowship (I.L.), 
NSF grant AST-1715145 (A.J.B.), a Kellett Mid-Career Award 
and a WARF Named Professorship from the University of Wisconsin-Madison 
Office of the Vice Chancellor for Research and Graduate Education with funding from the 
Wisconsin Alumni Research Foundation (A.J.B.), the William F. Vilas Estate (A.J.T.), a Wisconsin Space Grant Consortium Graduate and Professional Research Fellowship (A.J.T.), and a Sigma Xi Grant in Aid of Research (A.J.T.). 

The W.~M.~Keck Observatory is operated as a scientific partnership among the California Institute of Technology, the University of California, and NASA, and was made possible by the generous financial support of the W.~M.~Keck Foundation. 

We wish to recognize and acknowledge the very significant cultural role and reverence that the summit of Maunakea has always had within the indigenous Hawaiian community.  We are most fortunate to have the opportunity to conduct observations from this mountain.

\facilities{Keck:II (DEIMOS)}

\software{\cite{Astropy_1, Astropy_2}}

\bibliographystyle{aasjournal}
\bibliography{mybibliography.bib}

\appendix

\begin{figure*}[htb!]
\centering
\includegraphics[scale = 0.65]{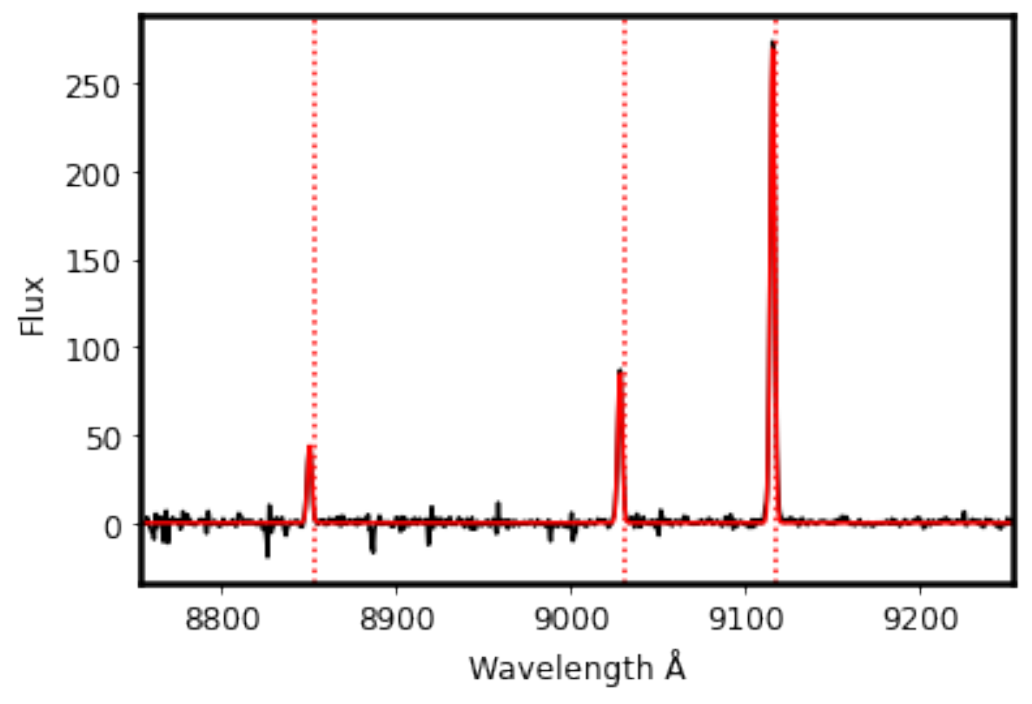} 
\includegraphics[scale = 0.65]{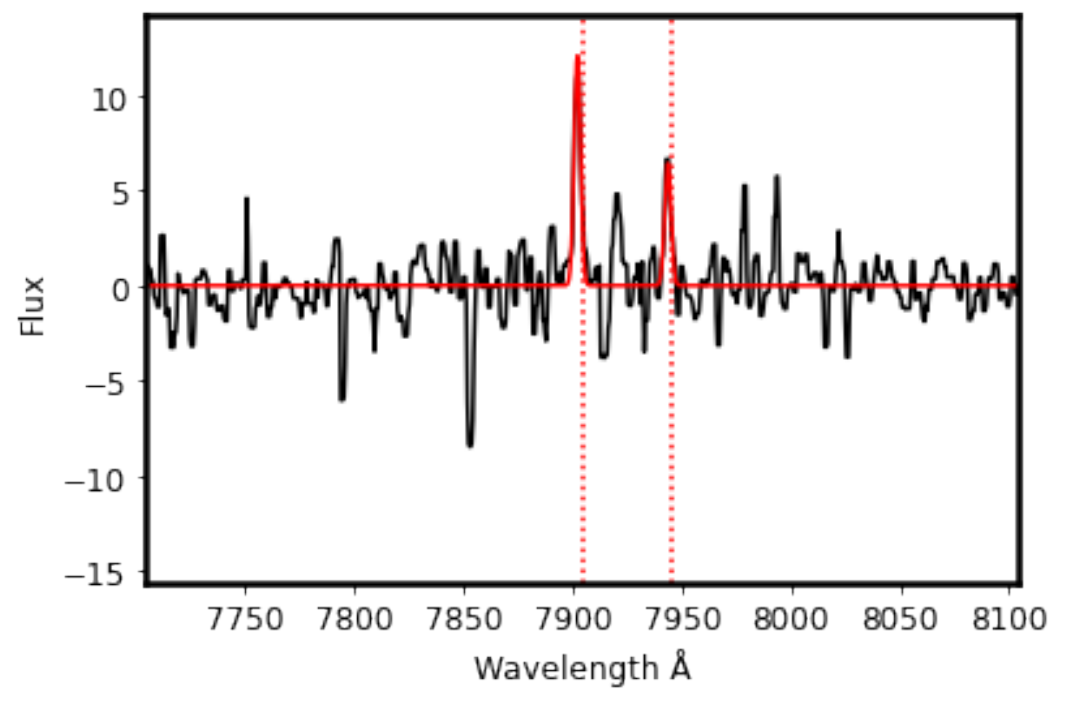}

\includegraphics[scale = 0.65]{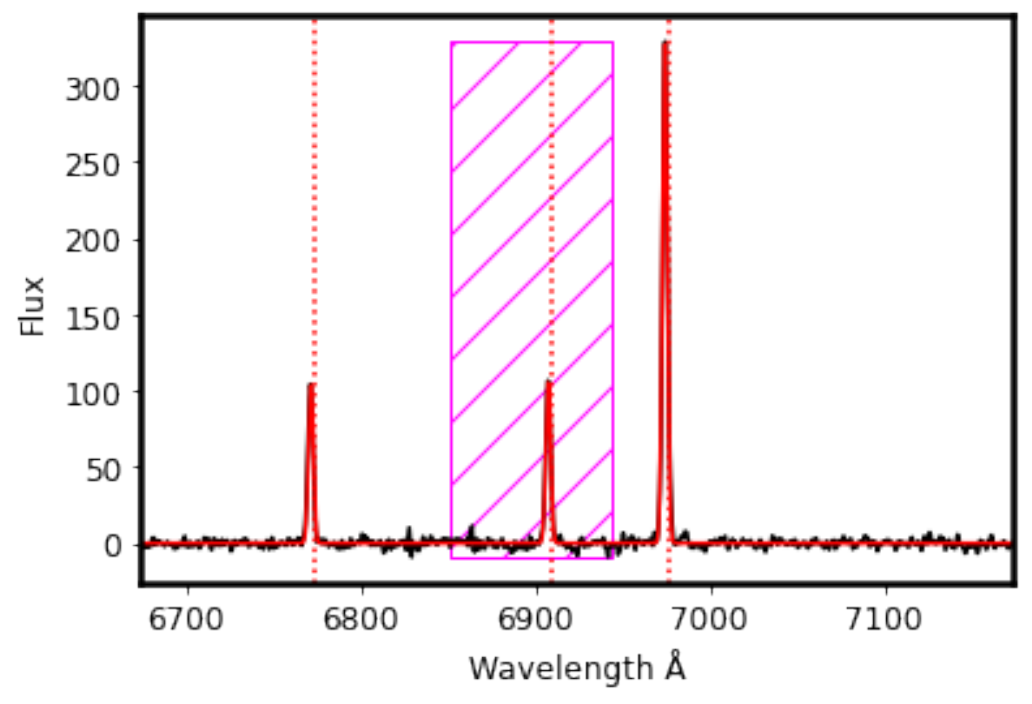} 
\includegraphics[scale = 0.65]{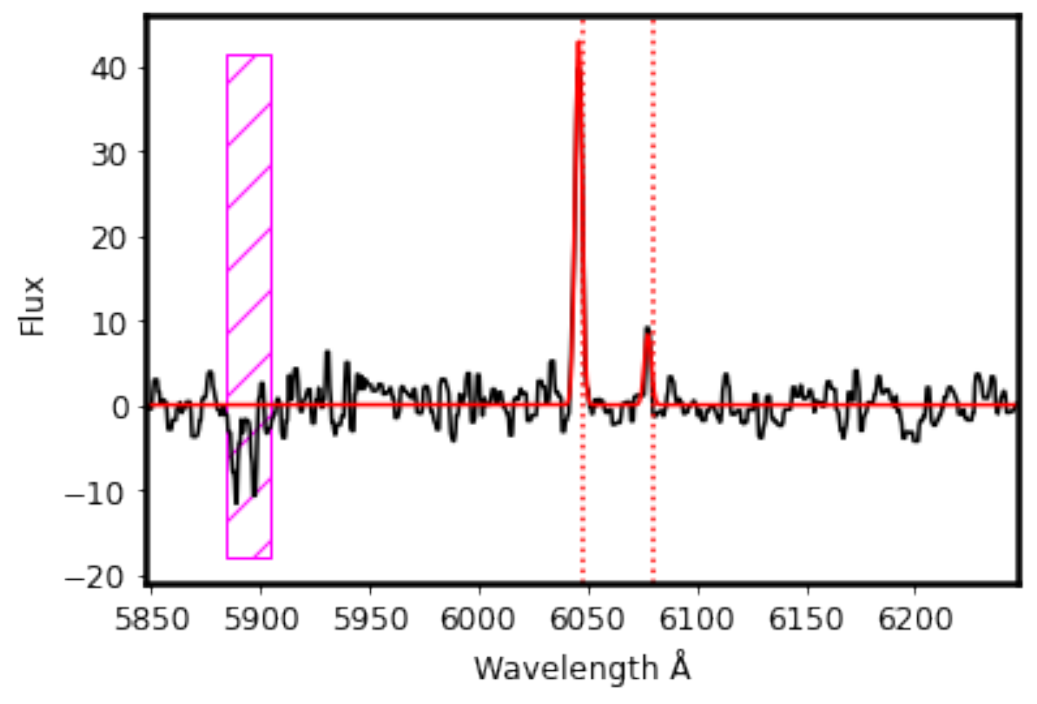}

\includegraphics[scale = 0.65]{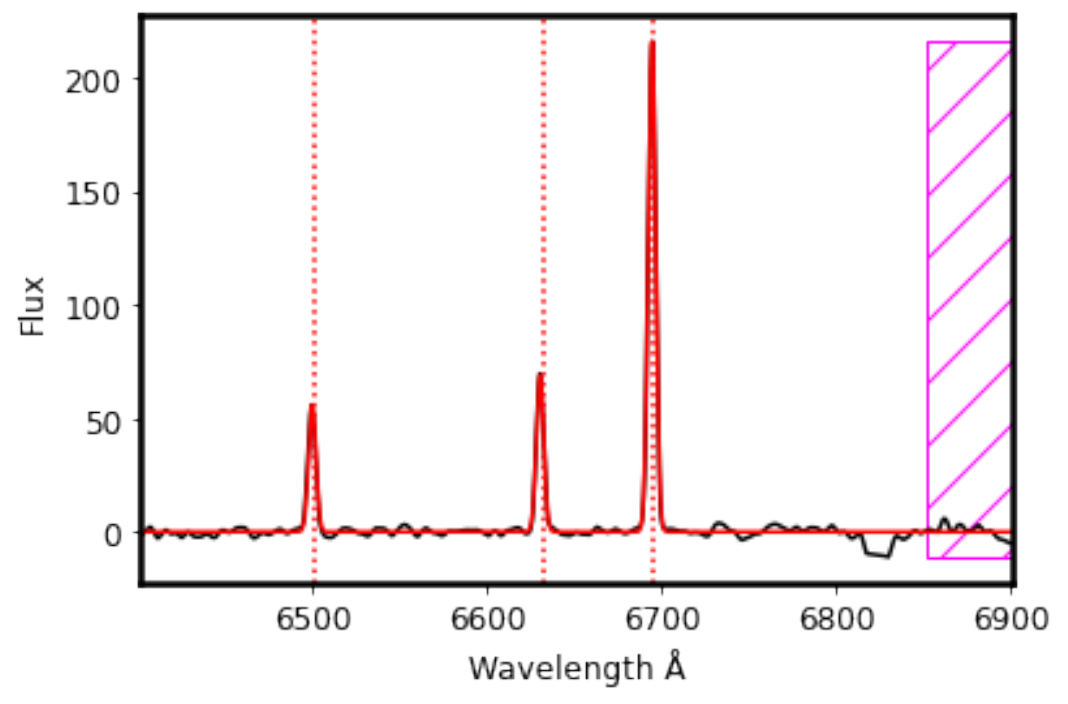} 
\includegraphics[scale = 0.65]{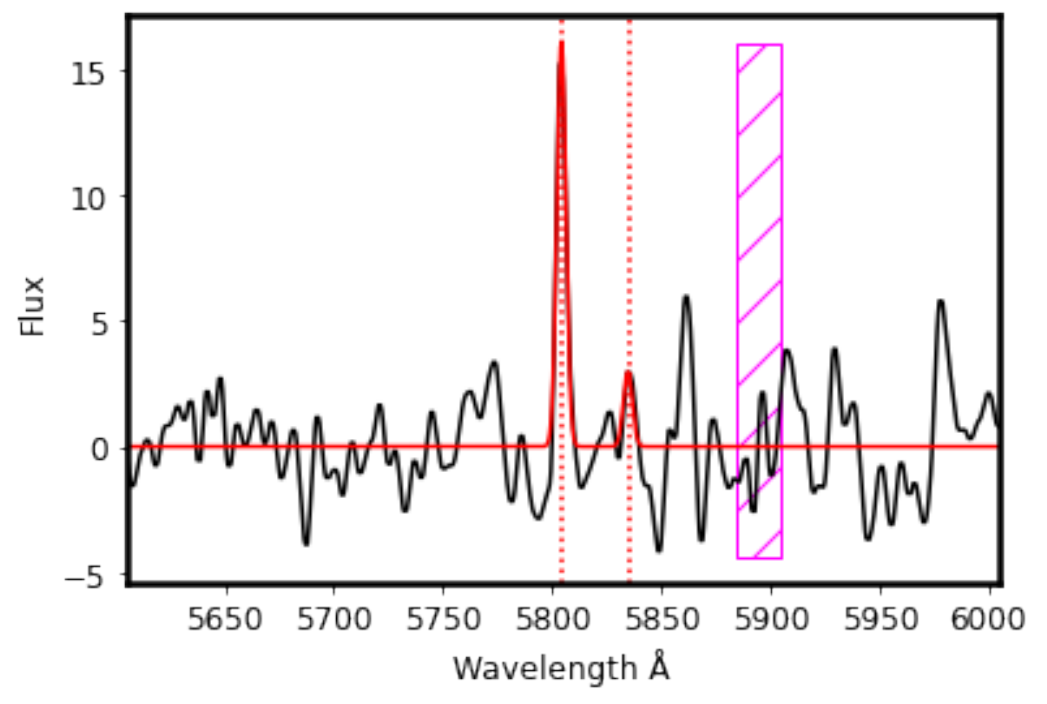}

\includegraphics[scale = 0.65]{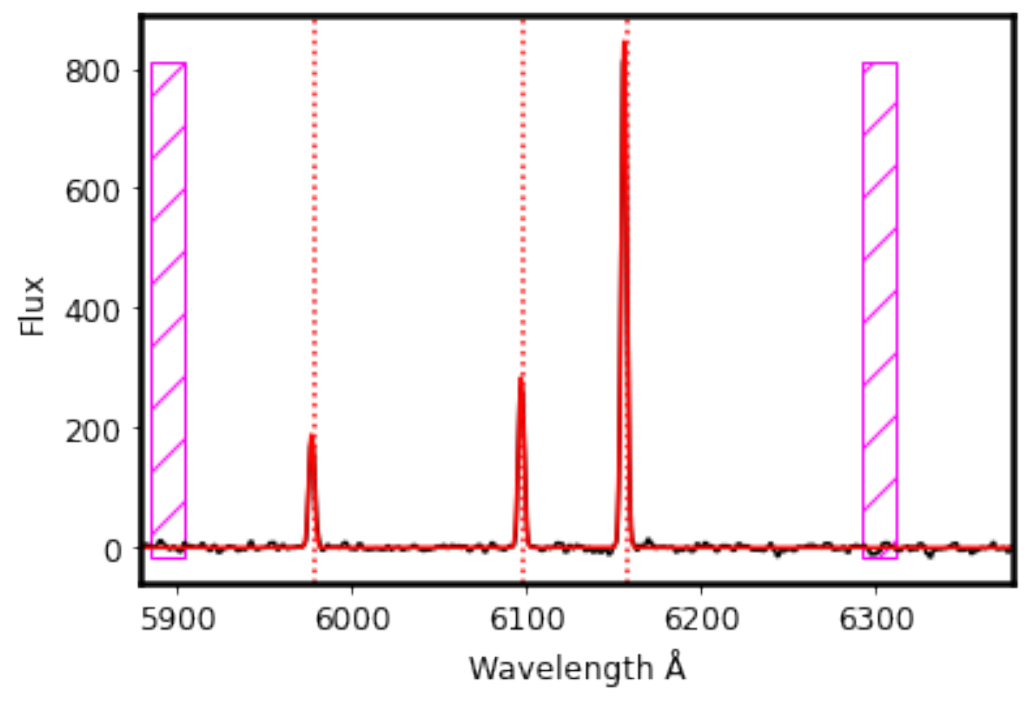} 
\includegraphics[scale = 0.65]{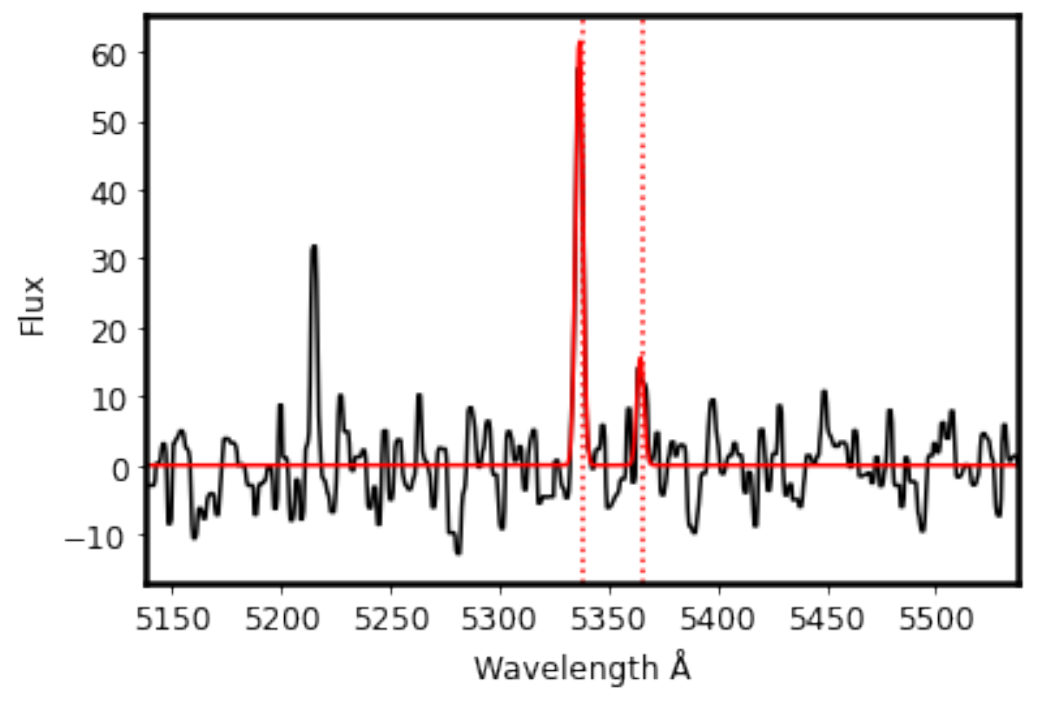}

\caption{\textit{Left}: Emission lines H$\beta$, [OIII]$\lambda 4959$, and [OIII]$\lambda 5007$ 
for our XMPGs with increasing metallicity, as presented in Table~\ref{table:only_table} and starting with the third source,
along with their triple Gaussian fits.
\textit{Right}: Emission lines H$\gamma$ and [OIII]$\lambda 4363$ for the same galaxies, 
along with their double Gaussian fits. A 5-point median smoothing 
has been applied. 
}

\end{figure*}

\begin{figure*}[htb!]
\setcounter{figure}{9}
\centering
\includegraphics[scale = 0.65]{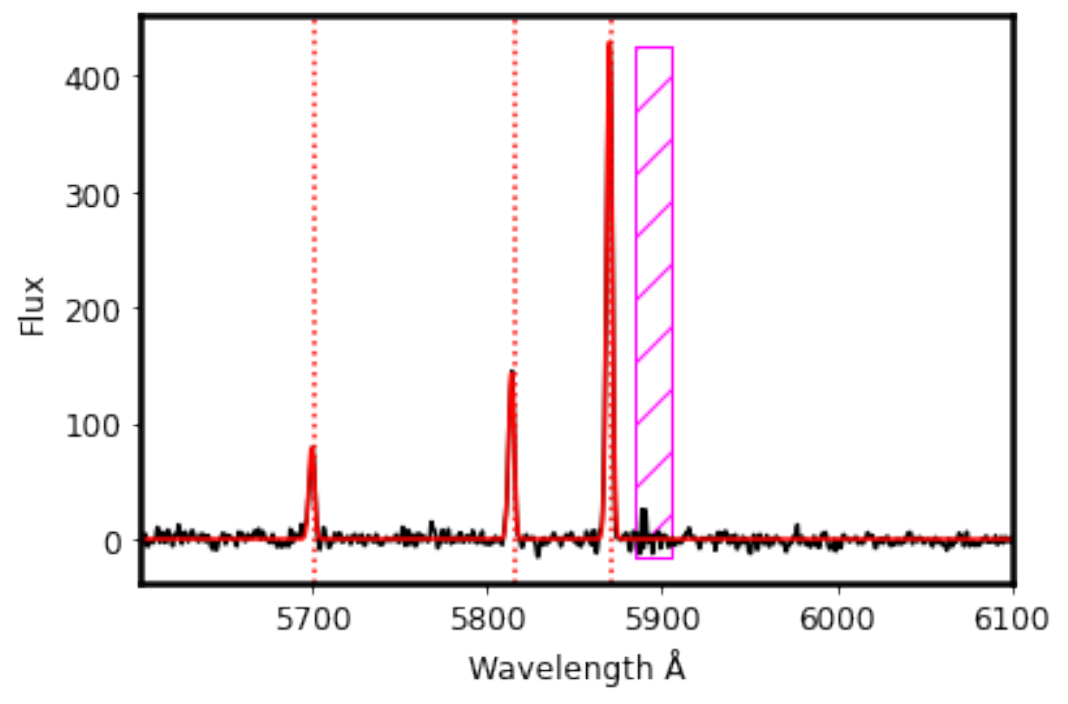} 
\includegraphics[scale = 0.65]{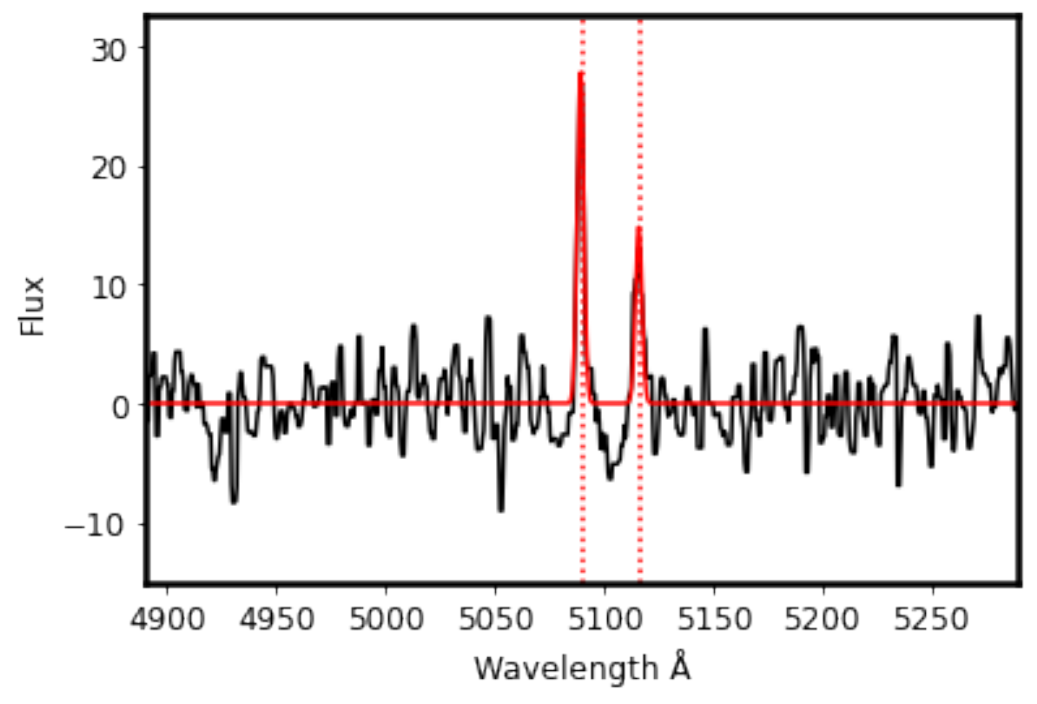}

\includegraphics[scale = 0.65]{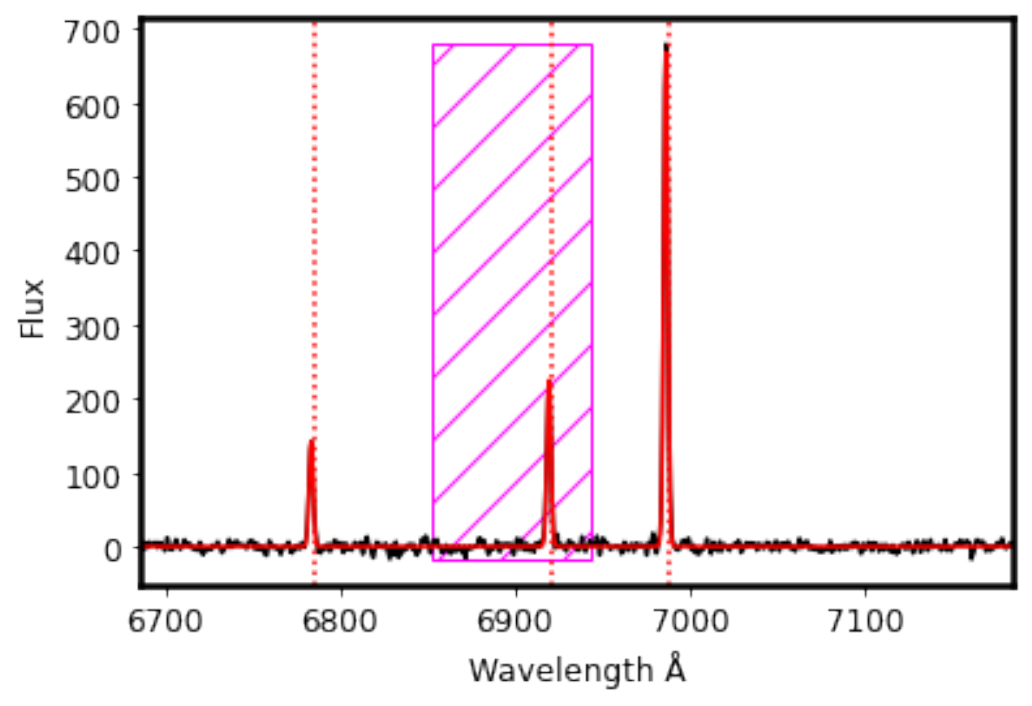} 
\includegraphics[scale = 0.65]{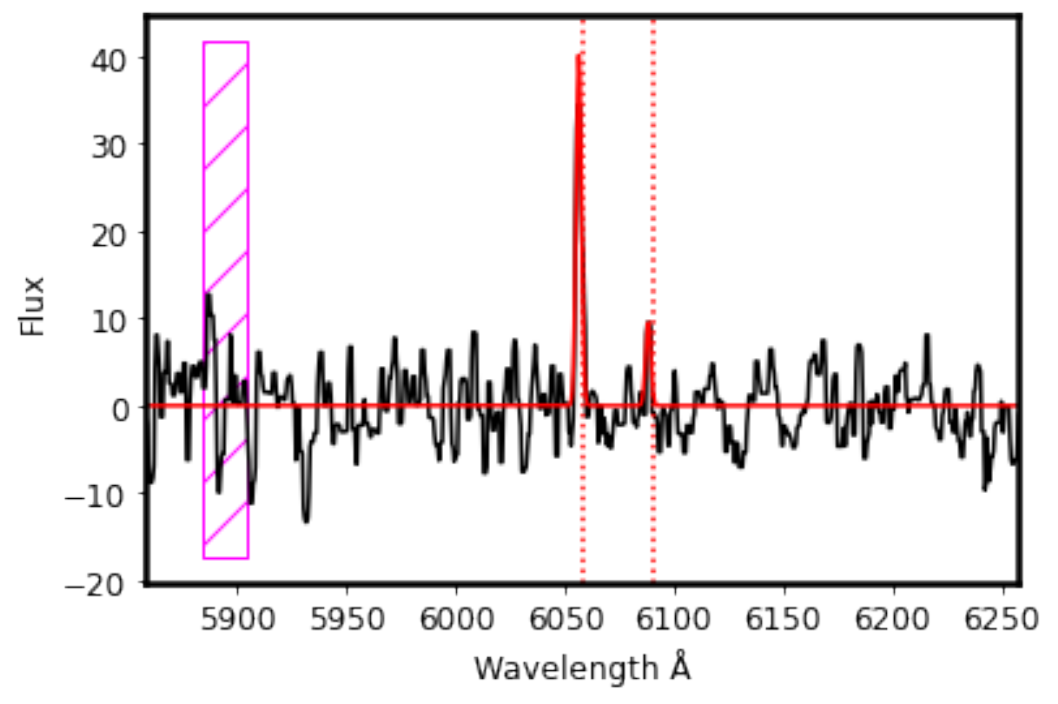}

\includegraphics[scale = 0.65]{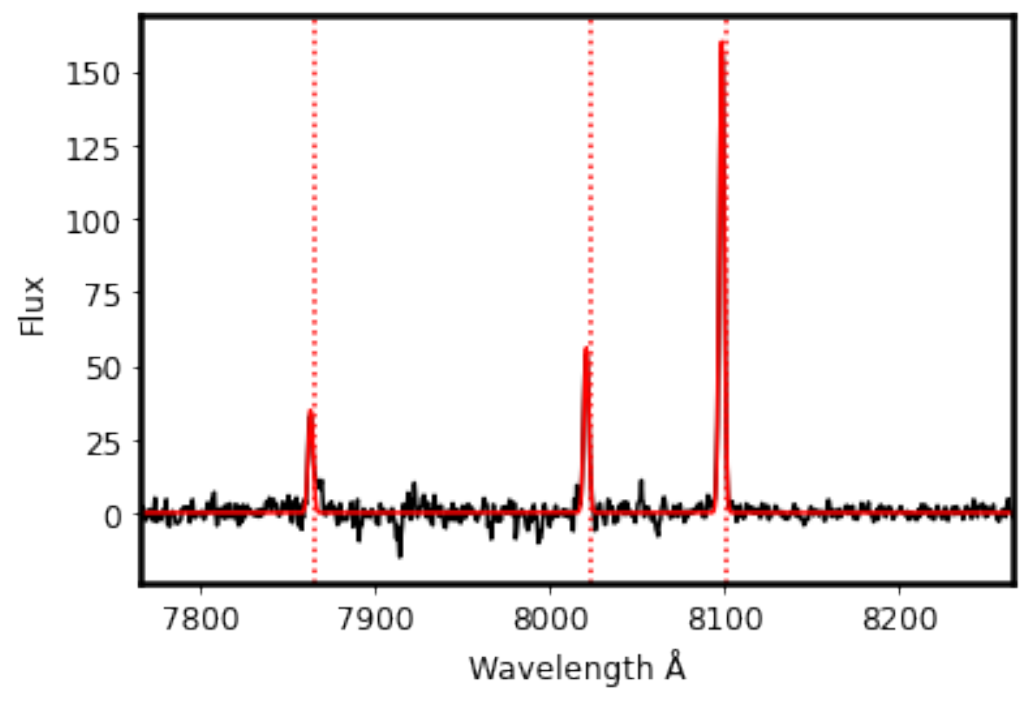} 
\includegraphics[scale = 0.65]{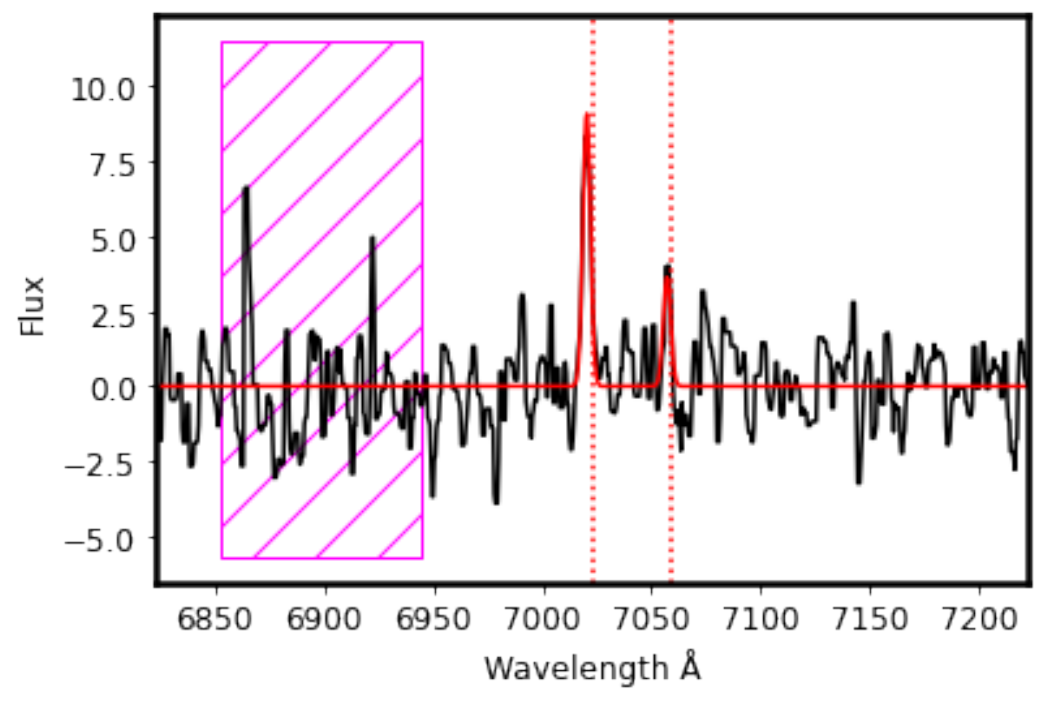}

\includegraphics[scale = 0.65]{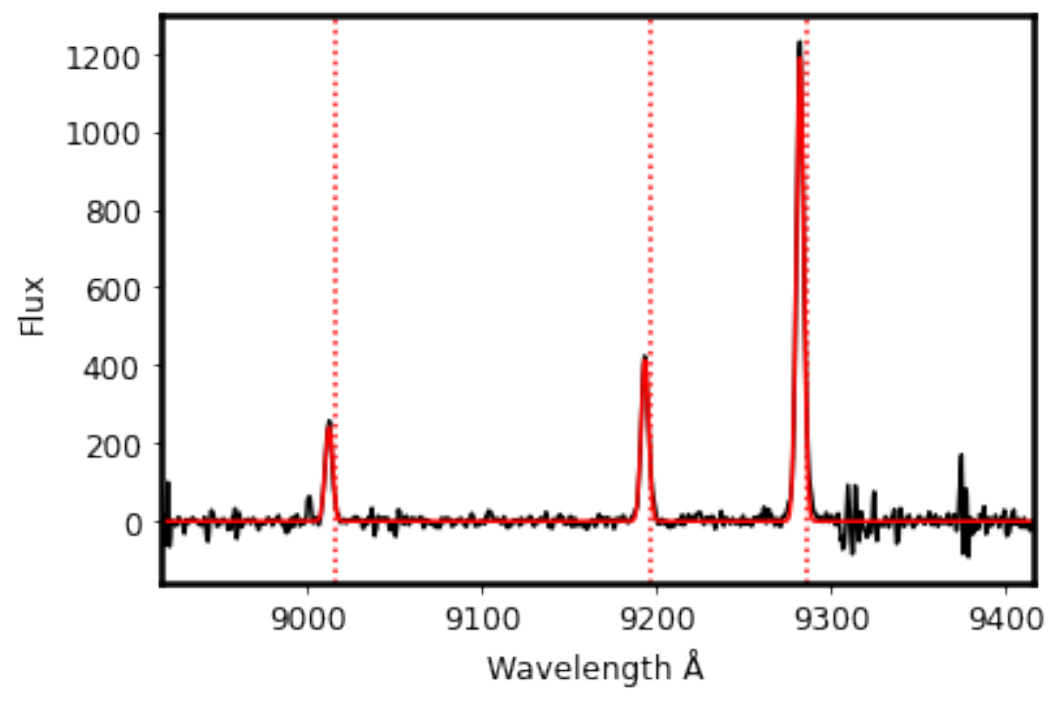} 
\includegraphics[scale = 0.65]{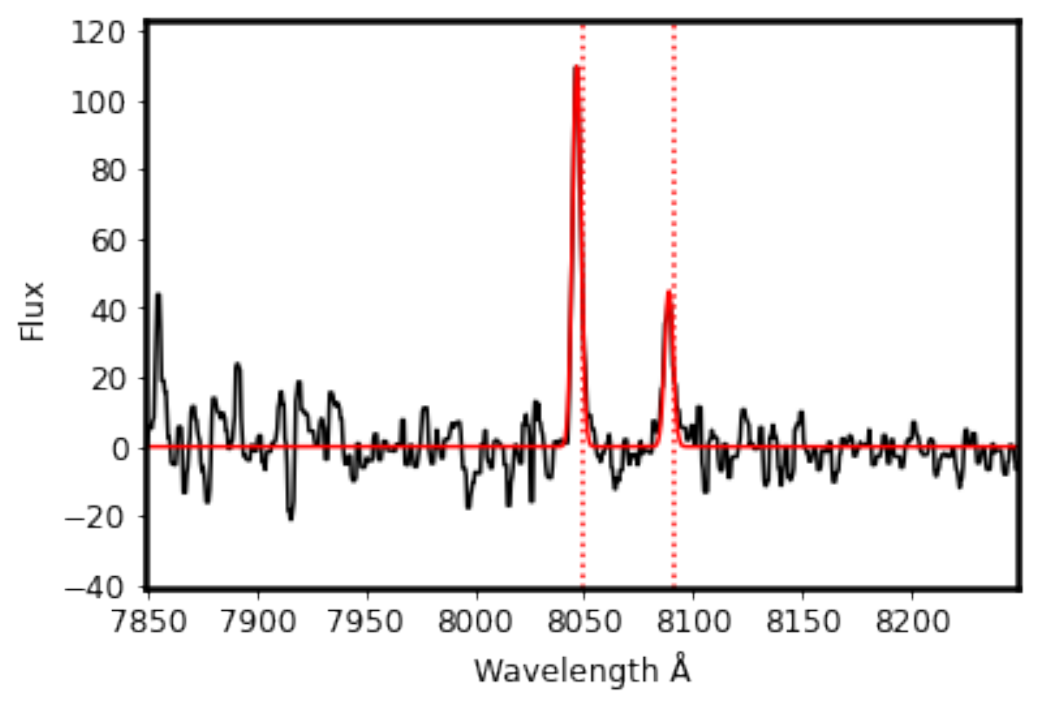}

\caption{(Cont).
}

\end{figure*}

\begin{figure*}[htb!]
\centering
\includegraphics[scale = 0.7]{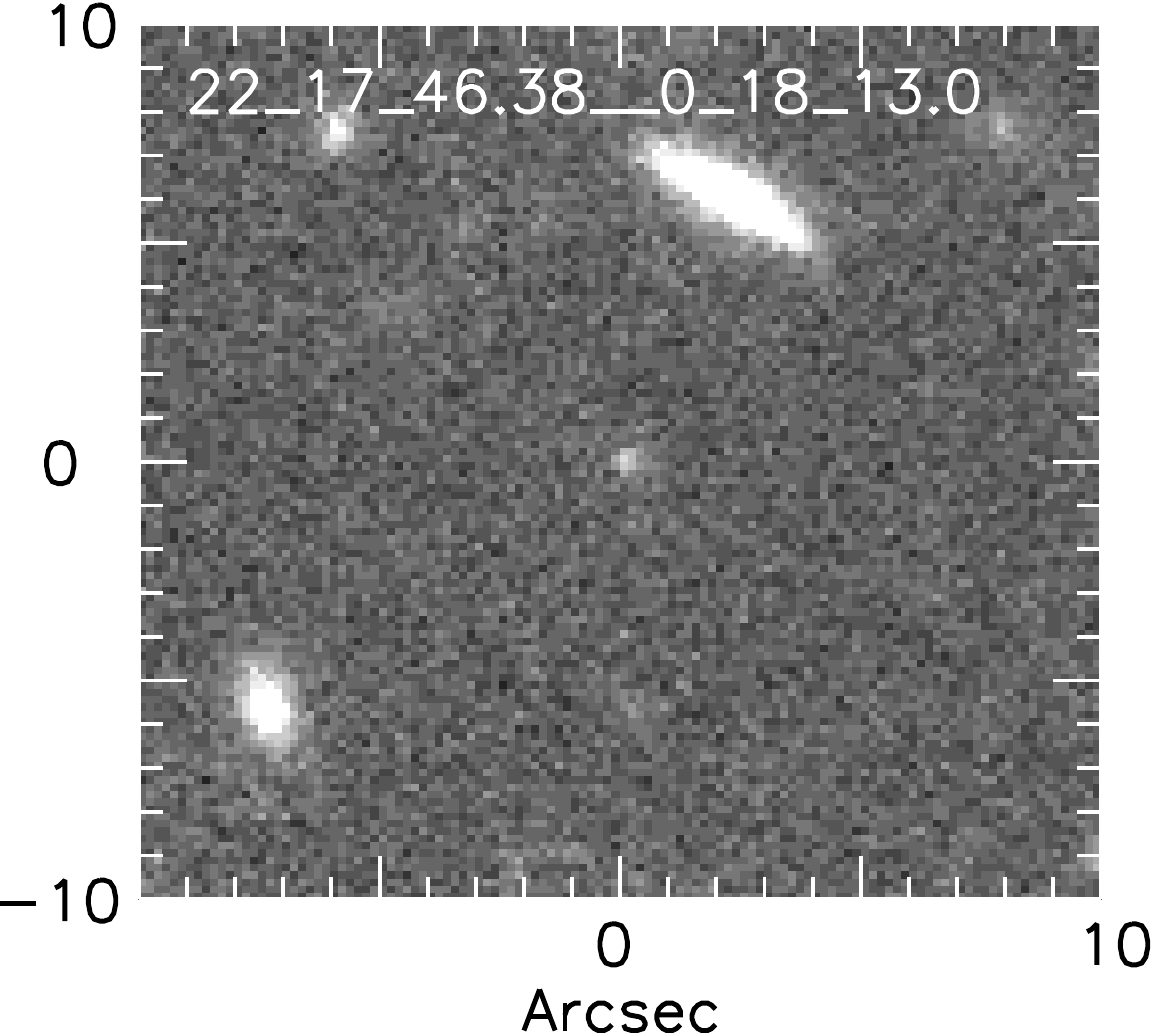} \includegraphics[scale = 0.7]{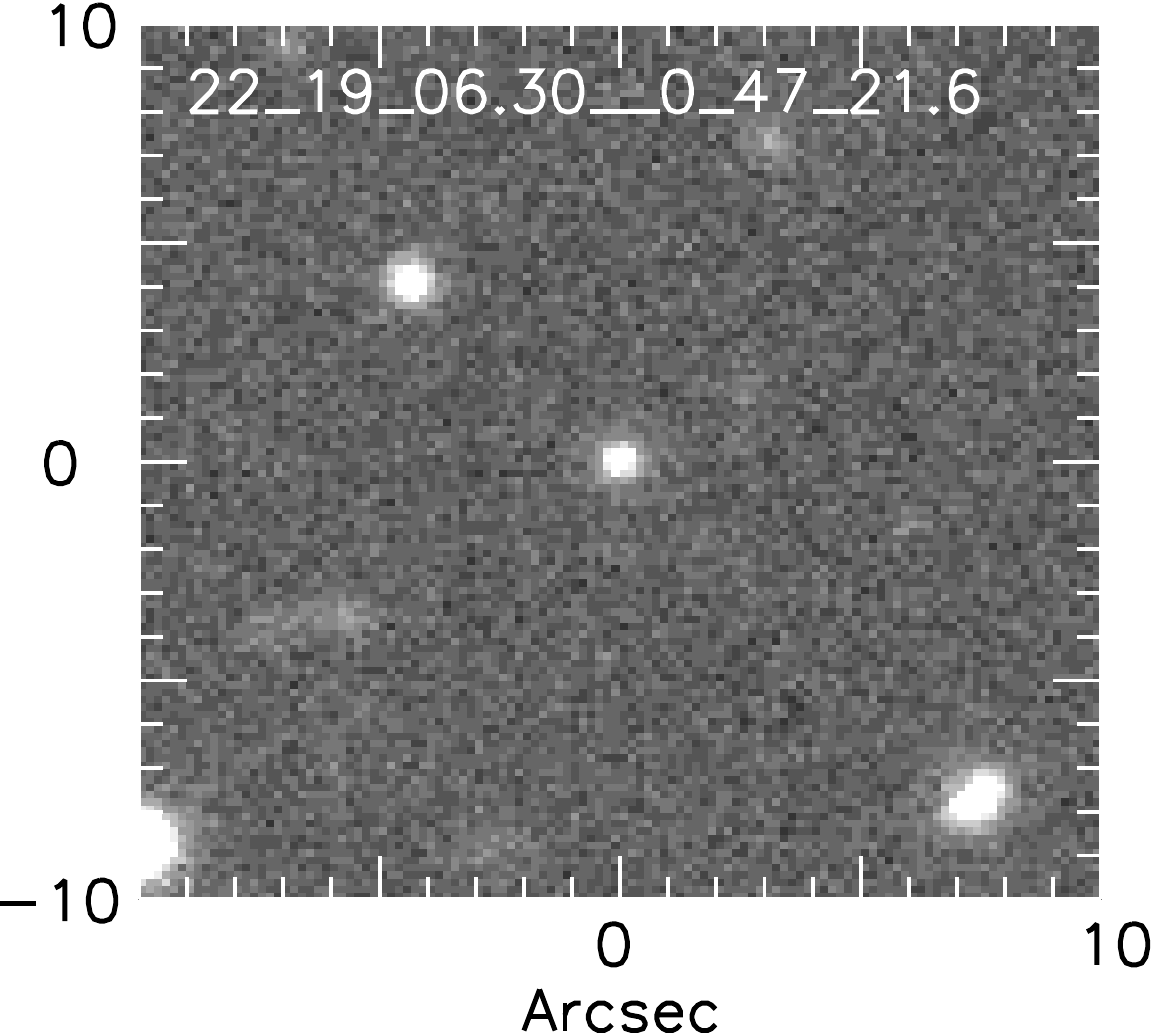}
\includegraphics[scale = 0.7]{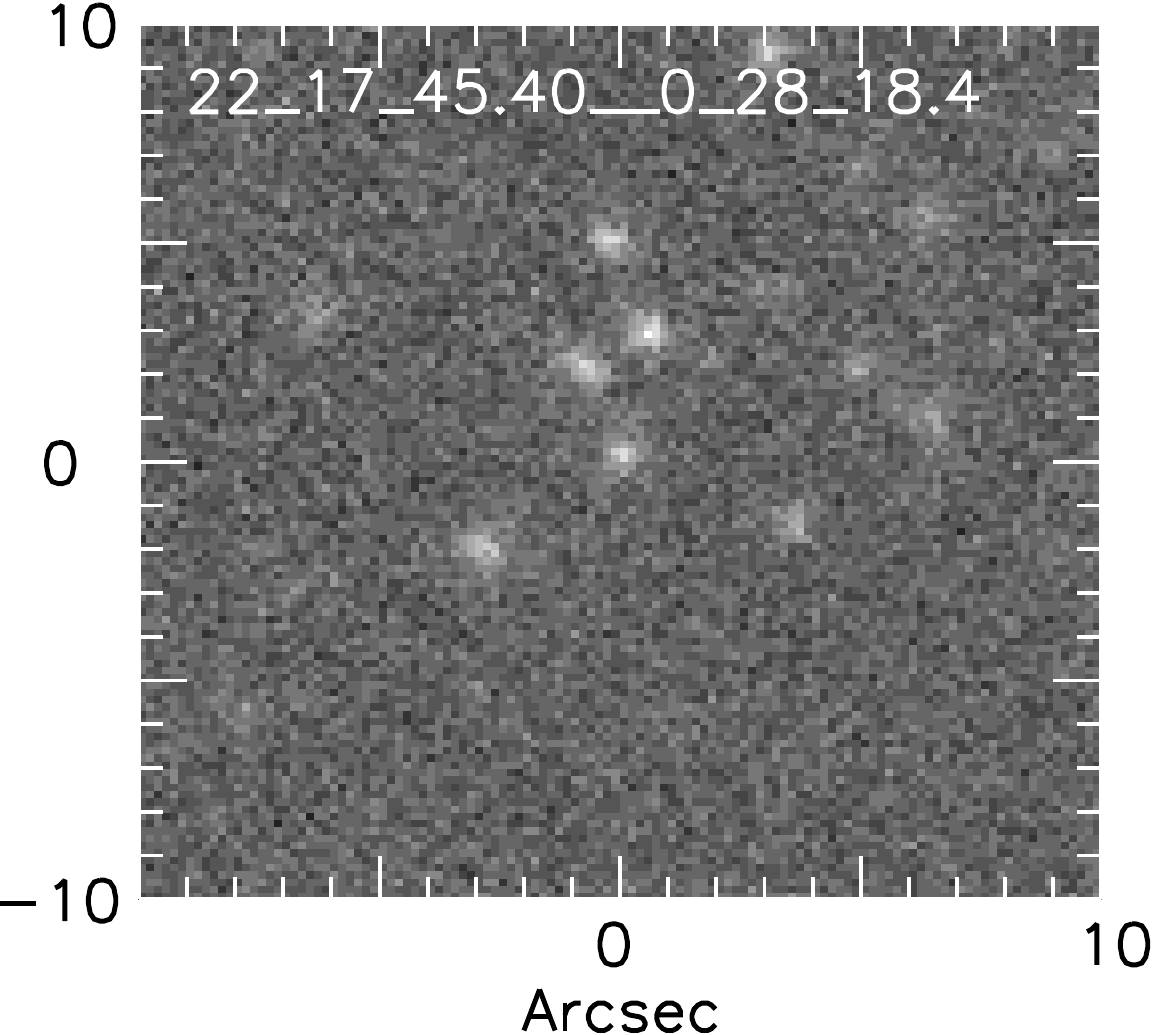}
\includegraphics[scale = 0.7]{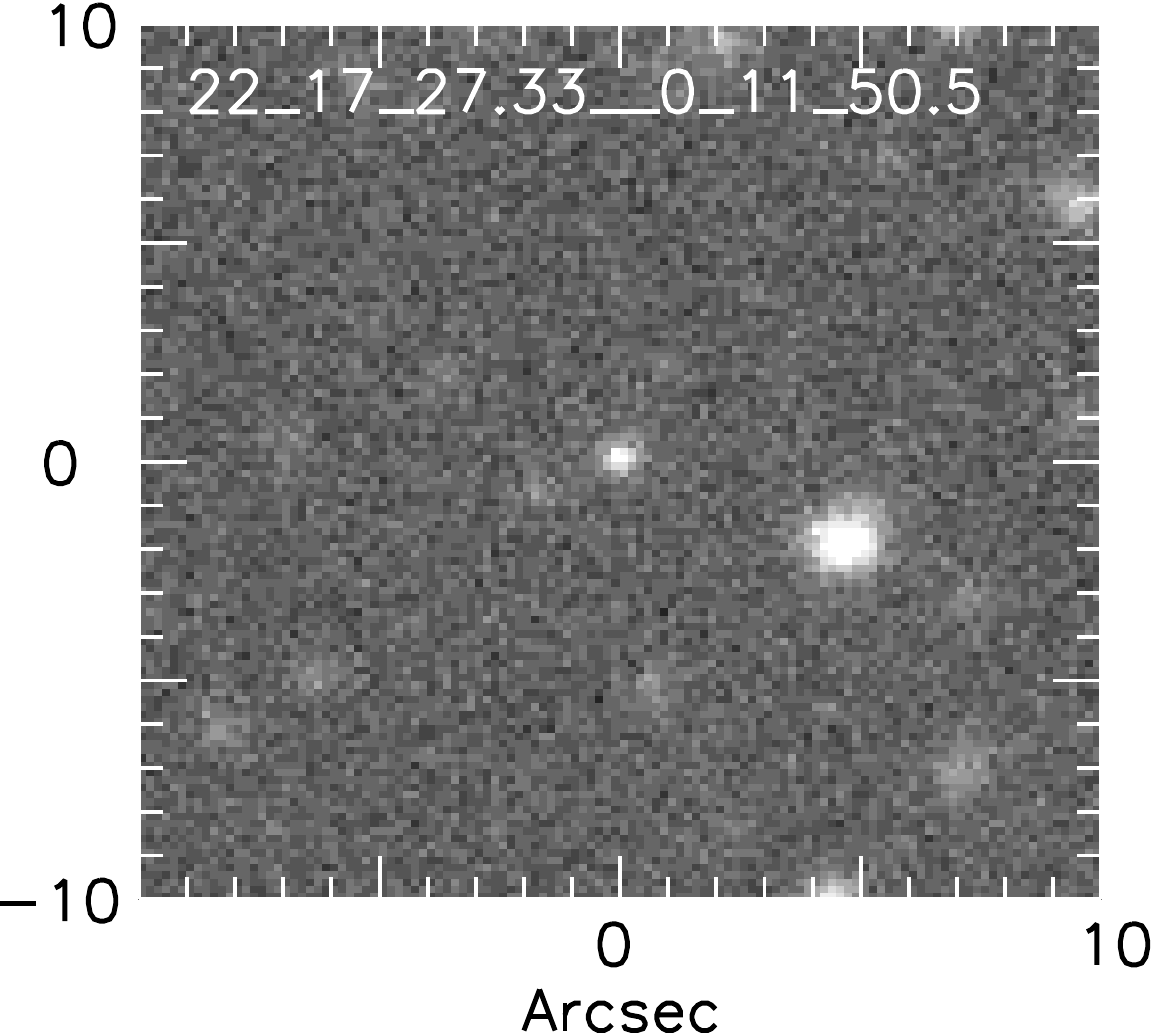}

\caption{$\rm 20\arcsec$ x $\rm 20\arcsec$ newly reduced Subaru Hyper Suprime-Cam z-band imaging of our four XMPGs in the SSA22 field (B. Radzom et al. 2022, in preparation). As expected with high quality imaging of XMPGs, we observe a sample of compact galaxies, which matches the morphology observed in other XMPG studies \citep[e.g.,][]{Izotov_2005, Izotov_2018, Hsyu_2018, Kojima_2020}.}

\end{figure*}

\end{document}